\documentclass[preprint,12pt,times]{elsarticle}
\usepackage{amssymb}
\usepackage{amsmath}
\usepackage{graphics}
\usepackage{graphicx}
\usepackage{subfigure}
\usepackage{dcolumn}
\usepackage{bm}
\usepackage{hyperref}

\journal{Journal of Non-Newtonian Fluid Mechanics}

\begin{document}

\begin{frontmatter}

\title{Correction of Wall Adhesion Effects in Batch Settling of Strong Colloidal Gels}

\author[CMIS]{D. R. Lester\fnref{label2}}
\ead{daniel.lester@csiro.au}
\fntext[label2]{ph: +61 3 9252 6195}
\author[MSACT,UOM]{R. Buscall}
\address[CMIS]{CSIRO Mathematics, Informatics and Statistics, PO Box 56, Highett,  Victoria 3190, Australia}
\address[MSACT]{MSACT Research and Consulting, Exeter, United Kingdom}
\address[UOM]{Particulate Fluids Processing Centre, Dept. of Chemical and Biomolecular Engineering, University of Melbourne, Victoria 3010, Australia}

\begin{abstract}
The batch settling test is widely used to estimate the compressive rheology of strongly flocculated colloidal suspensions, in particular the compressive yield strength and hydraulic permeability. Recently it has been discovered (D.R. Lester et al, J. Rheol., 58(5):1247–1276, 2014) that wall adhesion effects in these tests may be significantly greater than previously appreciated, which can introduce unbounded errors in the estimation of these rheological functions. Whilst a methodology to solve the underlying static problem and correct for wall adhesion effects has been developed, this method is quite complex and unwieldy, involving solution of a 2D hyper-elastic constitutive model for strong colloidal gels. In this paper we develop a highly simplified 1D visco-plastic approximation to the hyper-elastic model which admits analytic expressions for the equilibrium solids concentration profile and bed height. These expressions facilitate robust estimation of the compressive yield and wall adhesion strength via nonlinear regression of experimental data in the presence of small measurement errors.
\end{abstract}


\end{frontmatter}

\section{Introduction}

Solid-liquid separation of strongly flocculated colloidal gels is dictated by the compressive rheology of these materials, where the relevant rheological functions (compressive yield stress, hydraulic permeability) govern a wide range of separation processes, from sedimentation and continuous thickening through to centrifugation and pressure filtration. Central to characterization of compressive rheology over a range of solids volume fraction ($\phi$) is the batch settling test, the transient behaviour of which pertains to the hydraulic permeability $R(\phi)$, whilst equilibrium behaviour is dictated by the compressive yield stress $P_y(\phi)$. To extract these rheological functions, a one-dimensional (1D) vertical momentum balance is typically used~\citep{Kynch:52,Michaels/Bolger:62,HowellsEA:90,Landman/White:94,LesterEA:05,Diehl:07} to model sedimentation and consolidation of the suspension toward equilibrium, yielding an inverse problem regarding estimation of compressive rheology from experimental data.

Whilst several transient hydrodynamic mechanisms compromise the assumption of 1D settling behaviour, including inherent Rayleigh-Taylor instabilities within the suspension and velocity gradients arising from no-slip wall conditions, at equilibrium the major challenge arises from adhesion between colloidal particles and the container wall. This attractive force manifests as the wall adhesion (shear) strength $\tau_w(\phi)$, which, similar to the bulk shear yield strength, is a strongly increasing function of the solids concentration $\phi$. As equilibrium data (either in the form of solids volume fraction profiles or a series of bed heights) is used to determine the compressive yield stress, which in turn facilitates deconvolution of the hydraulic permeability from transient data, it is of paramount importance that this equilibrium state is modelled correctly.

Previous studies of the wall adhesion strength~\citep{Michaels/Bolger:62} showed that for the electrolytically coagulated colloidal suspensions tested, $\tau_w$ was small in comparison to the compressive yield strength $P_y(\phi)$, and so wall adhesion effects in wider columns are negligible, as reflected by the 1D equilibrium momentum balance (assuming $\phi$ is constant radially),
\begin{equation}
\frac{d P_y}{d\phi}\frac{d\phi_\infty}{dz}-\Delta\rho g\phi_\infty+\frac{2\tau_w(\phi)}{R}=0,\label{eqn:1Dbalance}
\end{equation}
where $z$ is the vertical bed depth (downwards from the suspension/supernatant interface), $\phi_\infty$ is the equilibrium solids concentration, $\Delta\rho$ the interphase density difference, $g$ gravitational acceleration constant, and $R$ is the radius of the batch settling column. As the apparent wall adhesion strength $\tau_w(\phi)$ typically appears~\citep{SethEA:08,BuscallEA:93,Barnes:95} to be of the same order as the bulk suspension shear yield strength $\tau_y(\phi)$, and the ratio of shear to compressive yield strength $S(\phi):=\tau_y(\phi)/P_y(\phi)$ appears to vary over the range 0.001-0.2~\citep{BuscallEA:87,BuscallEA:88,deKretserEA:02,ZhouEA:01,Channell/Zukoski:97}, the 1D approximation at equilibrium appears to be generally sound for all but narrow settling columns. Hence wall adhesion effects have been largely ignored over the past half century.

Recently, it has been established~\cite{LesterEA:14} that wall adhesion effects may be much more significant and prevalent than previously appreciated, especially for colloidal suspensions flocculated with high molecular-weight polymer flocculants which exhibit shear/compressive strength ratios of order 0.1. These suspensions, along with many coagulated systems, are both strongly cohesive and adhesive, and so readily adhere to settling column walls. For such materials it has been shown that impractically wide settling columns ($R\sim$ 1[m]) are required to render wall adhesion effects negligible, whilst for narrower columns, wall adhesion effects can introduce very serious errors in estimates of the compressive yield strength. When wall adhesion effects are significant the particulate network experiences a combination of shear and compressive stresses. From (\ref{eqn:1Dbalance}), the wall adhesion stress acts to counteract the interphase gravitational force up to the point that the entire suspension weight can be supported by wall shear stress alone. This state corresponds to the critical solids concentration $\phi_c$
\begin{equation}
\tau_w(\phi_c)\approx\frac{1}{2}\Delta\rho g \phi_c R,\label{eqn:phicritical}
\end{equation}
which represents an asymptotic limit for the equilibrium solids volume fraction profile $\phi_\infty(z)$. As such, the criterion upon which the relative magnitude of wall adhesion effects should be judged is not simply the ratio $S(\phi)$, but rather the magnitude of the wall adhesion strength $\tau_w(\phi)$ with respect to the wall shear stress $\frac{1}{2}\Delta\rho g \phi R$, where over a range of $\phi$ the former term must be much smaller than the latter for the 1D approximation to hold.

Hence wall adhesion effects can be highly significant for both polymer flocculated and electrolytically coagulated strong colloidal gels, and for other types of gel when the particle size is small, especially in the nano-scale range, since the strength ratio $S(\phi)$ is expected to increase with decreasing particle size. In general is it almost impossible to detect the presence or magnitude of wall adhesion errors from the equilibrium solids volume fraction profile alone. The only clear-cut case is that where the asymptotic limit $\lim_{z\rightarrow\infty}\phi_\infty(z)\rightarrow\phi_c$ is approached, and in this case errors arising from wall adhesion effects are unbounded as the corresponding predicted compressive yield stress diverges. As such, there is a clear and direct need to develop methods to detect and correct for wall adhesion effects in equilibrium batch settling data.

In a previous study \cite{LesterEA:14} we developed and validated a mathematical model of the suspension equilibrium stress state in the presence of wall adhesion, and developed error estimates for the uncorrected compressive yield strength. Analysis of the governing multidimensional force balance and suspension behaviour under arbitrary tensorial loadings raised fundamental questions regarding the constitutive modeling of strongly flocculated colloidal suspensions, particularly the relative validity and utility of both visco-plastic and visco-elastic rheological models for strong colloidal gels. A solution method was developed based upon the 2D static problem relating to the equilibrium stress balance which facilitated accurate estimates of the compressive and shear yield strengths from equilibrium solids volume fraction profiles. Whilst successful, this method is somewhat unwieldy in that it involves iterated numerical solution of the governing hyper-elastic constitutive model which is both computationally intensive and algorithmically complex.

What is required is a more simple and robust means of accurately estimating the compressive and shear yield strengths from equilibrium batch settling data, such that analysis of experimental data can be performed routinely. In this paper we develop such a method using a highly simplified 1D analytic solution for the equilibrium solids concentration profile based upon a visco-plastic constitutive model. The validity of this approximate solution is tested and justified against both the full hyper-elastic solution and experimental data, and methods for estimation of the compressive rheological functions from either solids volume fraction profile data or multiple equilibrium height data are developed and tested.

In the following Section we briefly review the equations governing multi-dimensional batch settling along with the visco-elastic and visco-plastic constitutive models for the the rheology of strong colloidal gels. In Section \ref{sec:1Dapprox} we develop and validate a simplified 1D model based on the visco-plastic constitutive model, and in Section \ref{sec:volfrac} this model is applied to equilibrium solids volume fraction profile data to validate the method. This method is then extended to equilibrium height data in Section \ref{sec:eqmheight}, and a sensitivity analysis is performed on this method prior to conclusions being made in Section \ref{sec:conclusions}.

\section{Batch Settling Modelling of Strong Colloidal Gels}\label{sec:models}

\subsection{Governing Equations}

Prior to the development of simplified models for solution of the equilibrium batch settling problem in Section 3, we briefly review governing equations and constitutive approaches. Although this study is focussed on the batch settling experiment under equilibrium conditions, the integral hyper-elastic constitutive model is couched in terms of the deformation history of the suspension, and so this model requires evolution of the initial condition $\phi=\phi_0<\phi_g$ (where $\phi_g$ is the suspension gel point) to the equilibrium state. Under the assumption that during sedimentation the bulk suspension velocity $\mathbf{q}:=\mathbf{v}_s\phi+\mathbf{v}_f(1-\phi)$ is zero (where $\mathbf{v}_s$, $\mathbf{v}_f$ respectively are the solid and fluid phase velocities), the inter-phase force balance over the particulate network~\cite{LesterEA:10} simplifies to
\begin{equation}
\frac{\partial\phi}{\partial t}+\nabla\cdot\frac{(1-\phi)^2}{R(\phi)}\left(\nabla\cdot\Sigma^N+\Delta\rho\mathbf{g}\phi\right)=0,\label{eqn:simpleevoln}
\end{equation}
where $\Sigma^N=-p_N\mathbf{I}+\boldsymbol\sigma_N$ is the network stress tensor, defined~\citep{Batchelor:77} as the difference between the total suspension stress $\Sigma$ and the fluid stress $\Sigma^f$:
\begin{equation}
\Sigma^N:=\Sigma-\Sigma^f.\label{eqn:networkstress}
\end{equation}
At long times the batch settling transient solution (\ref{eqn:simpleevoln}) converges to the equilibrium state $\phi_\infty$, where the solid and fluid fluxes decay to zero, yielding a balance between the network stress gradient and gravitational force as
\begin{equation}
\nabla\cdot\Sigma^N+\Delta\rho\mathbf{g}\phi_\infty=0.\label{eqn:eqmbalance}
\end{equation}
Given appropriate boundary conditions, this equation is capable of describing the batch settling experiment under equilibrium conditions in the presence of wall adhesion effects. To close both the transient (\ref{eqn:simpleevoln}) and equilibrium (\ref{eqn:eqmbalance}) equations, an appropriate constitutive model for strong colloidal gels is required for the network stress tensor $\Sigma^N$.

\subsection{Visco-plastic Constitutive Model}

Strong colloidal gels are mostly commonly modelled as visco-plastic materials (such as Bingham or Herschel-Bulkley models) as they are generally brittle in shear, with typical yield strains of the order 0.01-1\%. For the present purposes, the yield strain can be taken to mean the apparent yield strain, given by the ratio of yield stress to linear shear modulus. Similarly, compression is modelled as a self-limiting visco-plastic process in terms of the compressive yield stress $P_y(\phi)$. Whilst these idealized models only capture the gross rheology, they are simple and tractable, and allow the suspension network pressure $p_N$ and deviatoric stress $\tau_N=|\boldsymbol\sigma_N|$ to be quantified directly in terms of the compressive and shear yield strengths as
\begin{align}
&p_N\approx P_y(\phi),\,\,\frac{D_s\phi}{Dt}\geqslant 0,\label{eqn:Pcompress}\\
&\tau_N\approx\left(\frac{\tau_y(\phi)}{\dot\gamma}+\eta(\phi,\dot\gamma)\right)\dot\gamma\,\,\,\text{for}\,\,\tau_N\geqslant\tau_y(\phi),\label{eqn:taushear}
\end{align}
where $\frac{D_s}{Dt}$ is the material derivative with respect to the solid phase, where $\tau_N$ is the 2nd invariant of $\boldsymbol\sigma_N$, $\eta(\phi,\dot\gamma)$ is the apparent suspension viscosity (which is typically non-Newtonian), and $\dot\gamma$ is the rate of shear strain.

\subsection{Hyper-elastic Constitutive Model}

However, as recent studies~\cite{SprakelEA:11,LindstomEA:12,GibaudEA:10,Santos:13,Koumakis:11,GibaudEA:08,Ovarlez:13,RamosEA:01,Ovarlez:07,CloitreEA:00,Tindley:07,Kumar:12,Uhlherr:05,GrenardEA:13} have established, the true nature of yield of these materials is more somewhat more complicated than is captured by visco-plastic constitutive models, and rather strong colloidal gels act as visco-elastic materials which undergo rapid strain hardening in compression and can undergo rapid strain softening and/or hardening in shear. As such, a more consistent and accurate way to describe strong colloidal gels is as visco-elastic media with strongly nonlinear shear modulus $G$ and bulk modulus $K$. Under this hyper-elastic formulation, the network pressure $p_N$ and deviatoric stress $\boldsymbol\sigma_N$ are given by the integral evolution equations~\cite{LesterEA:14}
\begin{align}
&p_N=\int_{\phi_0}^{\phi(t)}K(\phi)d\ln\phi,\label{eqn:hyperpressure}\\
&\boldsymbol\sigma_N=\int_{-\infty}^t \frac{\partial G(\phi,\gamma,t-s)}{\partial s}\boldsymbol\Phi(s) ds\label{eqn:hypershear},
\end{align}
where $\boldsymbol\Phi$ is the deviatoric component of the Hencky strain tensor $\mathbf{H}$ for the solid phase. Due to the integral nature of the hyper-elastic constitutive model, the initial state $\phi(\mathbf{x},0)=\phi_0$ must be evolved via the transient equation (\ref{eqn:simpleevoln}) toward the equilibrium state $\phi_\infty(\mathbf{x})$. Conversely, the visco-plastic model does not require the full deformation history, and so acts directly on the equilibrium force balance (\ref{eqn:eqmbalance}).

Hence the visco-plastic constitutive model represents an simplified approximation to the full hyper-elastic constitutive model which is less computationally intensive, but it also suffers from the drawback that the stress-strain relationship is ill-defined prior to yield. In the case of multi-dimensional batch settling, this corresponds to a statically indeterminate equilibrium stress state which requires special treatment to resolve.

\subsection{Closure of Viscoplastic Model}

To resolve this statically indeterminate state, the assumption of negligible normal stress differences
\begin{align}
N_1&:=\Sigma^N_{zz}-\Sigma^N_{rr}=0,\\
N_2&:=\Sigma^N_{rr}-\Sigma^N_{\theta\theta}=0,
\end{align}
was invoked~\cite{LesterEA:14} to close the equilibrium stress state, and it was found that under this closure the visco-plastic solution closely approximates the full hyperelastic solution. Whilst this closure may not be appropriate in general, i.e. strong colloidal gels may exhibit significant normal stress differences in other applications, these materials under static equilibrium do not exhibit such effects. In conjunction with this closure, the assumption of axisymmetry within a cylindrical batch settling column simplifies the network stress tensor to
\begin{equation}
\Sigma^N=\left(
           \begin{array}{ccc}
             \Sigma^N_{rr} & \Sigma^N_{r\theta} & \Sigma^N_{rz} \\
             \Sigma^N_{r\theta} & \Sigma^N_{\theta\theta} & \Sigma^N_{\theta z} \\
             \Sigma^N_{rz} & \Sigma^N_{\theta z} & \Sigma^N_{zz} \\
           \end{array}
         \right)
         =\left(
           \begin{array}{ccc}
             p_N & 0 & \tau_N \\
             0 & p_N & 0 \\
             \tau_N & 0 & p_N \\
           \end{array}
         \right),
\end{equation}
where $p_N=\frac{1}{3}\text{tr}(\Sigma^N)$, $\tau_N=|\boldsymbol\sigma^N)|=|\Sigma^N_{rz}|$ are the isotropic and deviatoric network stresses respectively. Under these simplifications, and ignoring the angular components $\theta$, the equilibrium network force balance reduces to
\begin{align}
\frac{\partial p_N}{\partial r}+\frac{\partial \tau_N}{\partial z}&=0,\label{eqn:radial}\\
\frac{1}{r}\frac{\partial}{\partial r}\left(r \tau_N\right)+\frac{\partial p_N}{\partial z}+\Delta\rho g \phi_\infty&=0,\label{eqn:vertical}
\end{align}
which applies over the suspension domain $\mathcal{D}:\{r,z\}\in[0,R]\times[0,h_\infty)$, where $h_\infty$ is the equilibrium suspension height. If the initial solids volume fraction $\phi_0$ is less than the suspension gel point $\phi_g$, then at equilibrium the network pressure $p_N$ is everywhere equal to the compressive yield strength
\begin{equation}
p_N(\mathbf{x})=P_y(\phi(\mathbf{x})_\infty)\forall\mathbf{x}\in\mathcal{D}.\label{eqn:phyphi}
\end{equation}
The hyperbolic system (\ref{eqn:radial}), (\ref{eqn:vertical}) is also subject to the boundary conditions
\begin{align}
p_N|_{z=0}&=0,\label{eqn:pressurez0}\\
\tau_N|_{z=0}&=0,\label{eqn:shearz0}\\
\tau_N|_{r=0}&=0,\label{eqn:shearr0}\\
\tau_N|_{r=R}&=\tau_y(\phi|_{r=R}),\label{eqn:shearrR}
\end{align}
where the critical boundary condition (\ref{eqn:shearrR}) may be recast via (\ref{eqn:phyphi}) as
\begin{equation}
\tau_N|_{r=R}=F_1(p_N|_{r=R}),
\end{equation} and the function $F_1$ reflects the relationship between the compressive yield strength $P_y(\phi)$ and shear yield strength $\tau_y(\phi)$ at a given volume fraction $\phi$.

Based upon physical arguments and experimental data, \citep{Buscall:09} proposes a relationship for the ratio $S(\phi)=\tau_y(\phi)/P_y(\phi)$ between the shear and compressive yield strength of a particulate gel, which rapidly decreases from around 1 near the gel point $\phi_g$ to the asymptotic value $S_\infty$ with increasing $\phi$. For a compressive yield strength function of the form
\begin{equation}
P_y(\phi)=k\left(\left(\frac{\phi}{\phi_g}\right)^n-1\right),\label{eqn:pyphi_func}
\end{equation}
the asymptotic value is
\begin{equation}
S_\infty=\kappa n \gamma_c.\label{eqn:Sinf}
\end{equation}
where $\gamma_c$ is the critical shear strain and $\kappa$ the ratio of shear to compressive moduli, which is related to the the Poisson ratio $\nu$ as
\begin{equation}
\kappa=\frac{2}{3}\left(\frac{1-\nu}{1-2\nu}\right),
\end{equation}
where $\nu=3/8$, $\kappa=5/3$ for systems bound by central forces as per Cauchy's relationships. For a compressive yield strength with the functional form (\ref{eqn:pyphi_func}), $S(\phi)$ is then
\begin{equation}
S(\phi)=\left(\left(\frac{1}{S_{\infty}}-1\right)\left(1-\left(\frac{\phi}{\phi_g}\right)^{-n}\right)+1\right)^{-1},\label{eqn:Sphi}
\end{equation}
and $F_1$ in the wall boundary condition (\ref{eqn:shearrR}) simplifies to
\begin{align}
F_1(p_N)=\frac{p_N}{\left(\frac{1}{S_\infty}+1\right)\left(\frac{p_N}{p_N+k}\right)+1}.\label{eqn:FpN}
\end{align}

\begin{figure}[h]
\centering
\begin{tabular}{c c c}
\includegraphics[width=0.32\columnwidth]{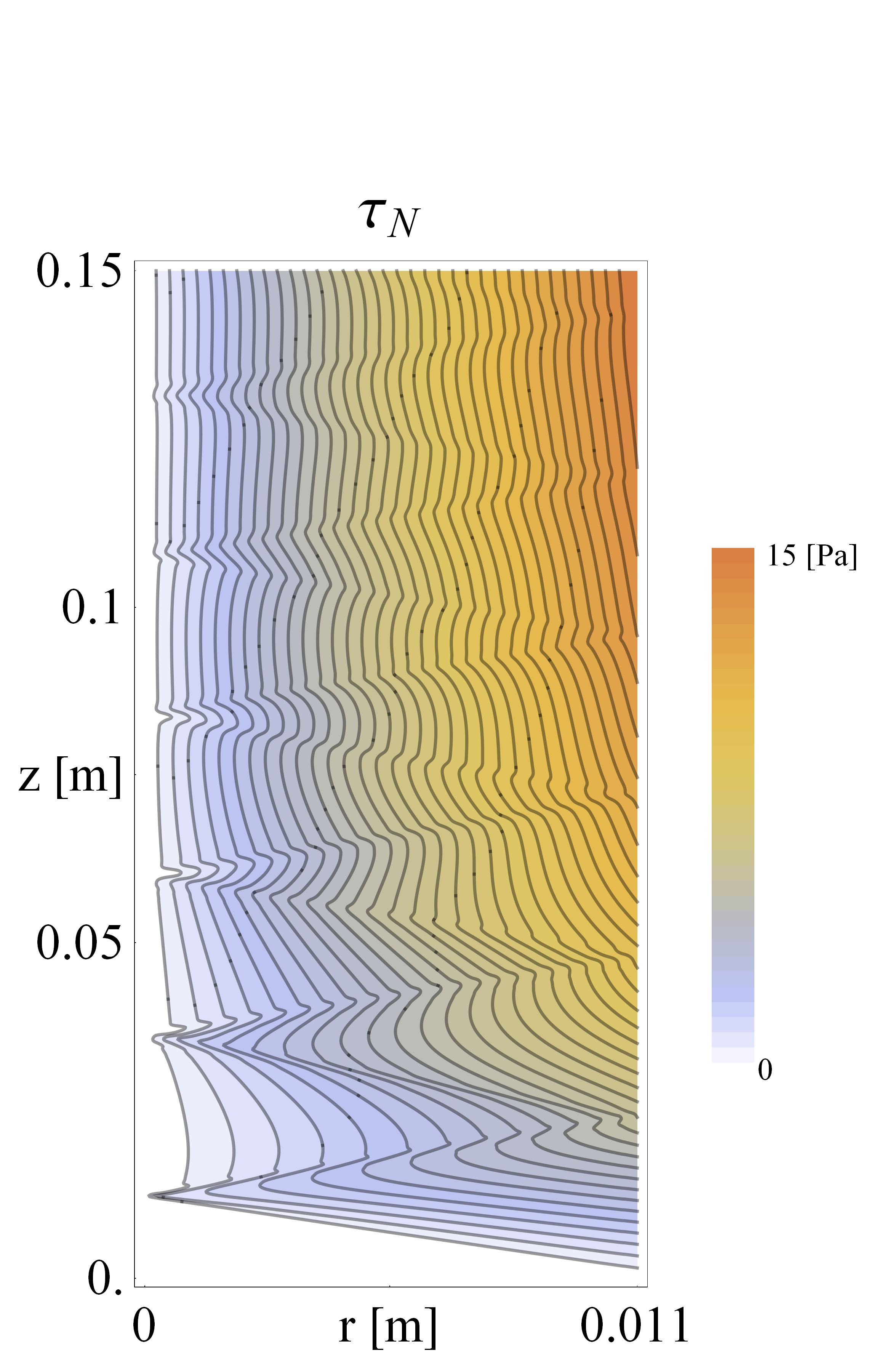}&
\includegraphics[width=0.32\columnwidth]{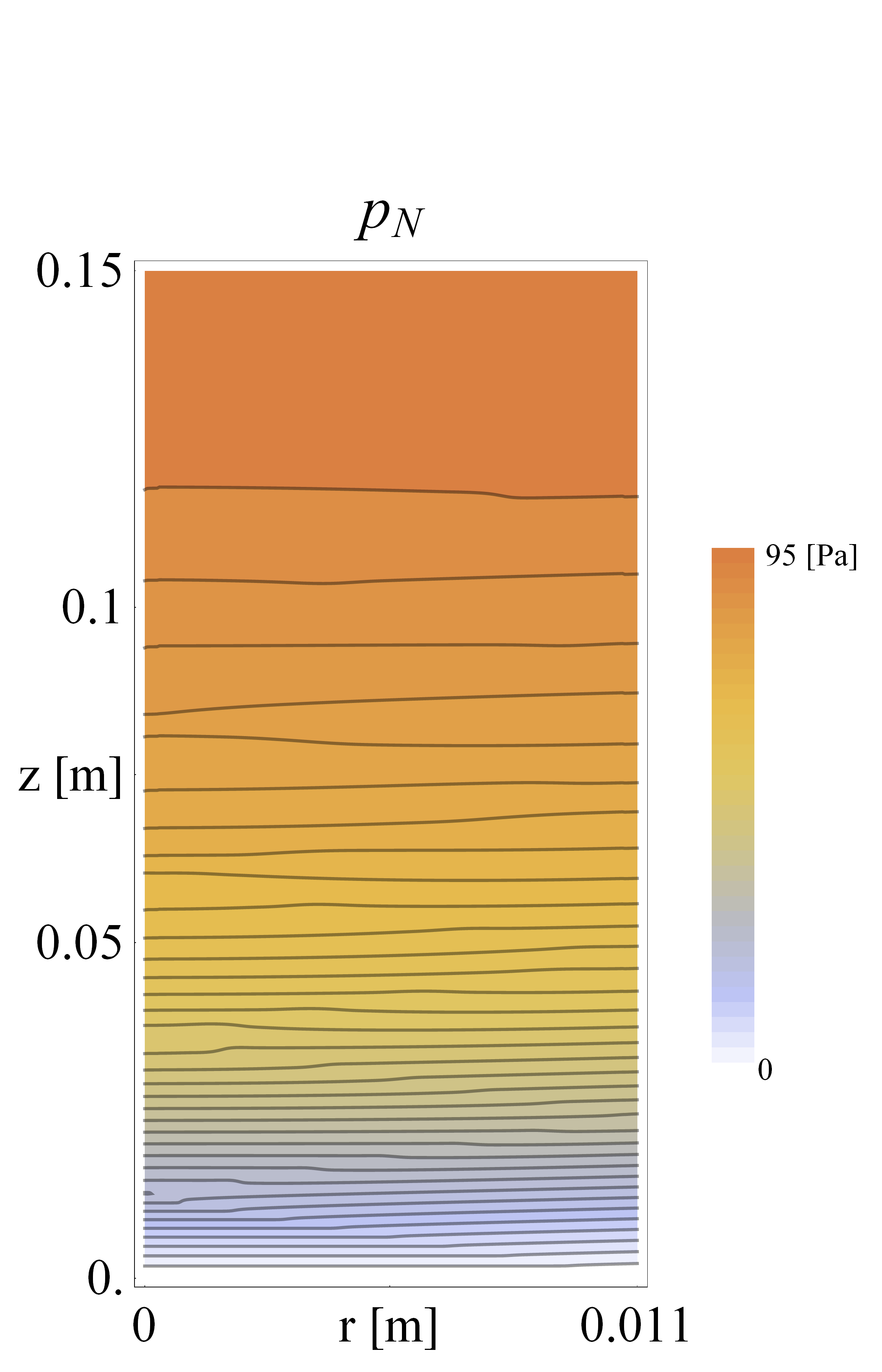}&
\includegraphics[width=0.32\columnwidth]{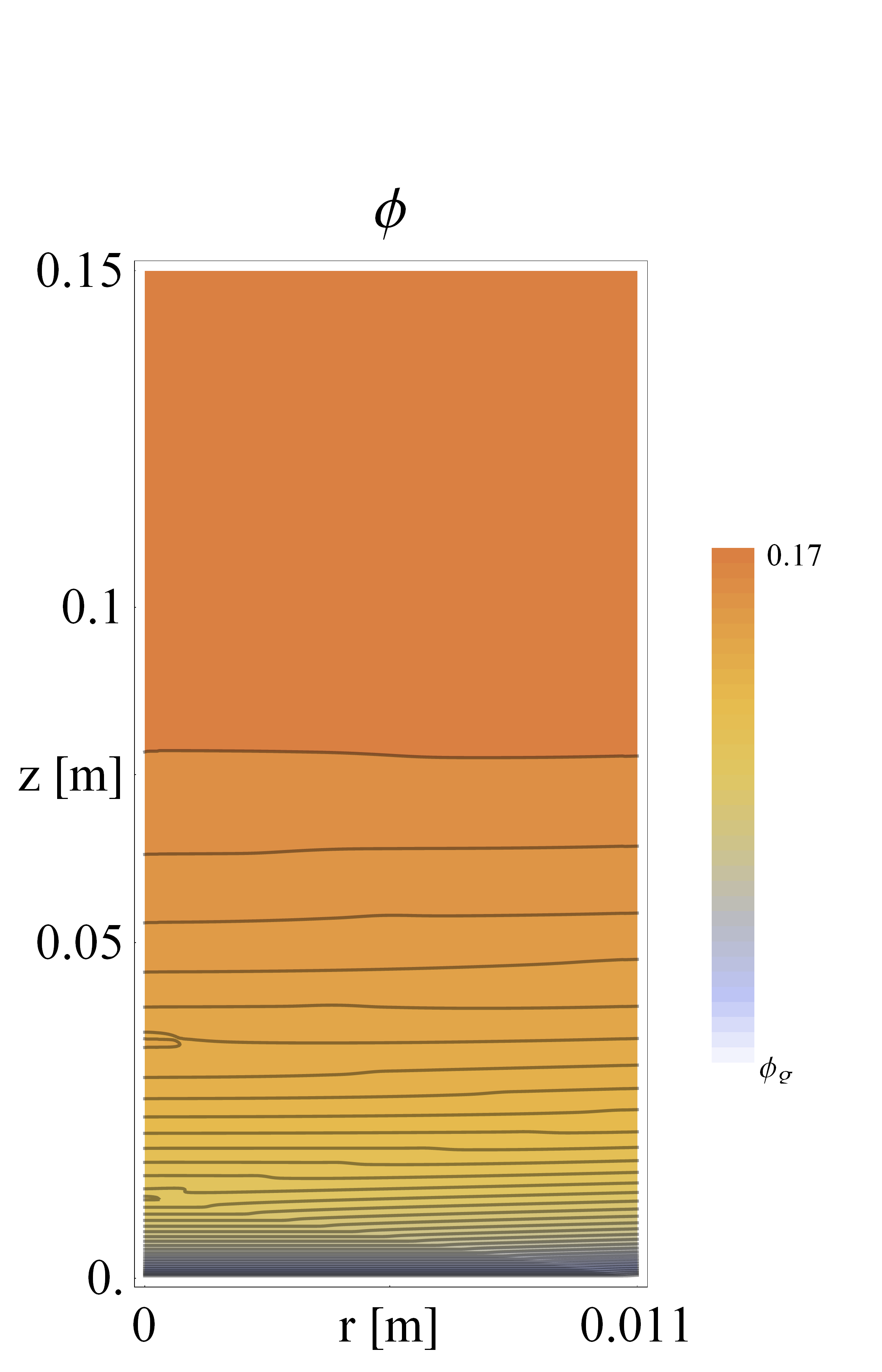}\\
(a) & (b) & (c)
\end{tabular}
\caption{Comparison of (a) network shear stress $\tau_N$, (b) network pressure $p_N$ and solids volume fraction $\phi_\infty$ distributions for visco-plastic constitutive model.}\label{fig:viscoplastic}
\end{figure}

The hyperbolic system (\ref{eqn:radial}) (\ref{eqn:vertical}) manifests as a pair of characteristics which propagate down the bed at approximately 45 degrees to the container wall and are reflected off the container wall and symmetry axis as per the reflection conditions (\ref{eqn:shearr0}), (\ref{eqn:shearrR}). In terms of the shear stress distribution, these characteristics manifest as $C^1$ shocks (i.e. non Lifshitz-continuous) in the shear stress distribution, as shown in Fig~\ref{fig:viscoplastic}(a), which are clearly non-physical. These artifacts are a direct result of the hyperbolic nature of the visco-plastic constitutive model, and call into question its validity for multi-dimensional static problems.


\begin{figure}[h]
\centering
\begin{tabular}{c c c}
\includegraphics[width=0.32\columnwidth]{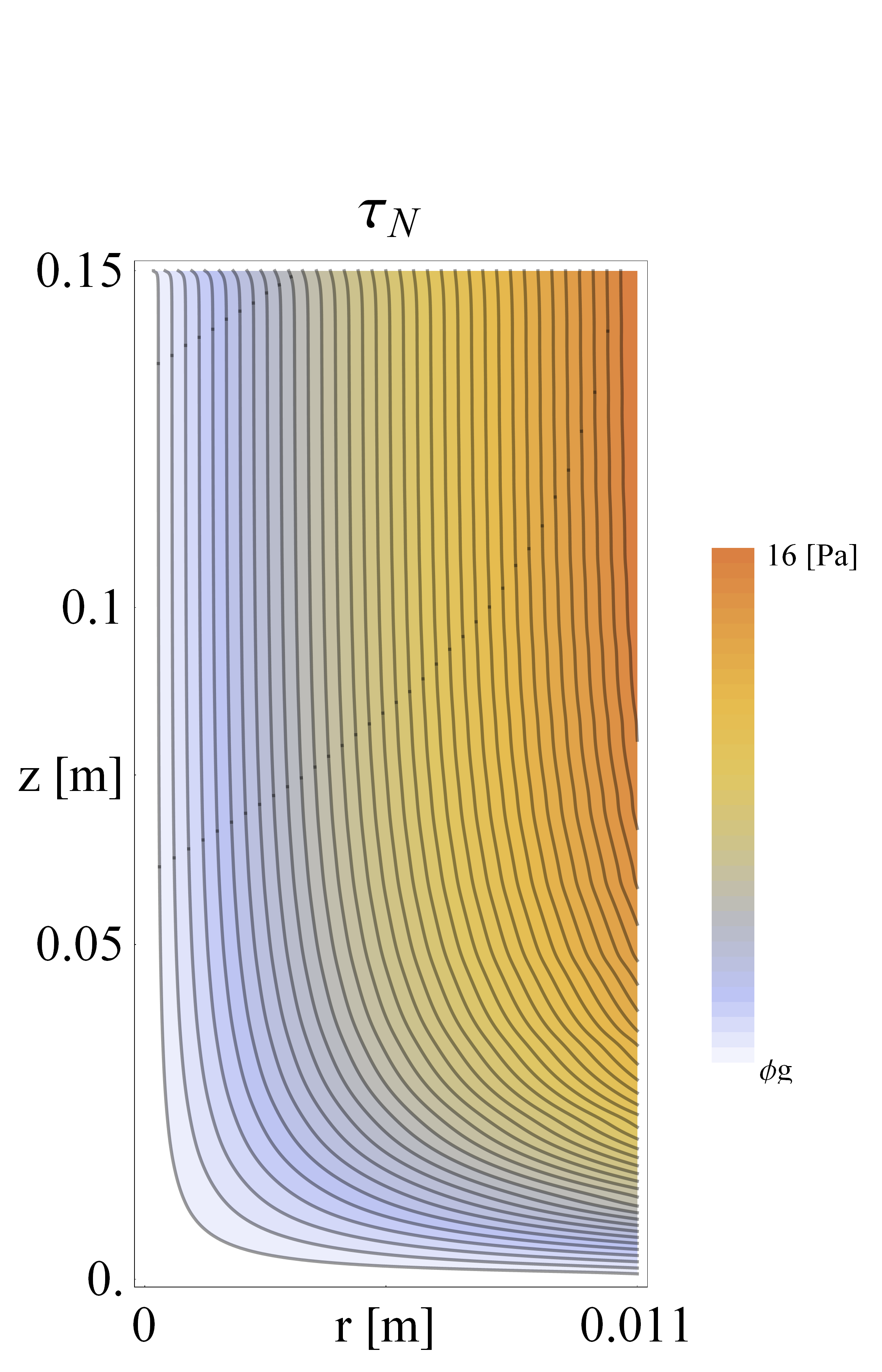}&
\includegraphics[width=0.32\columnwidth]{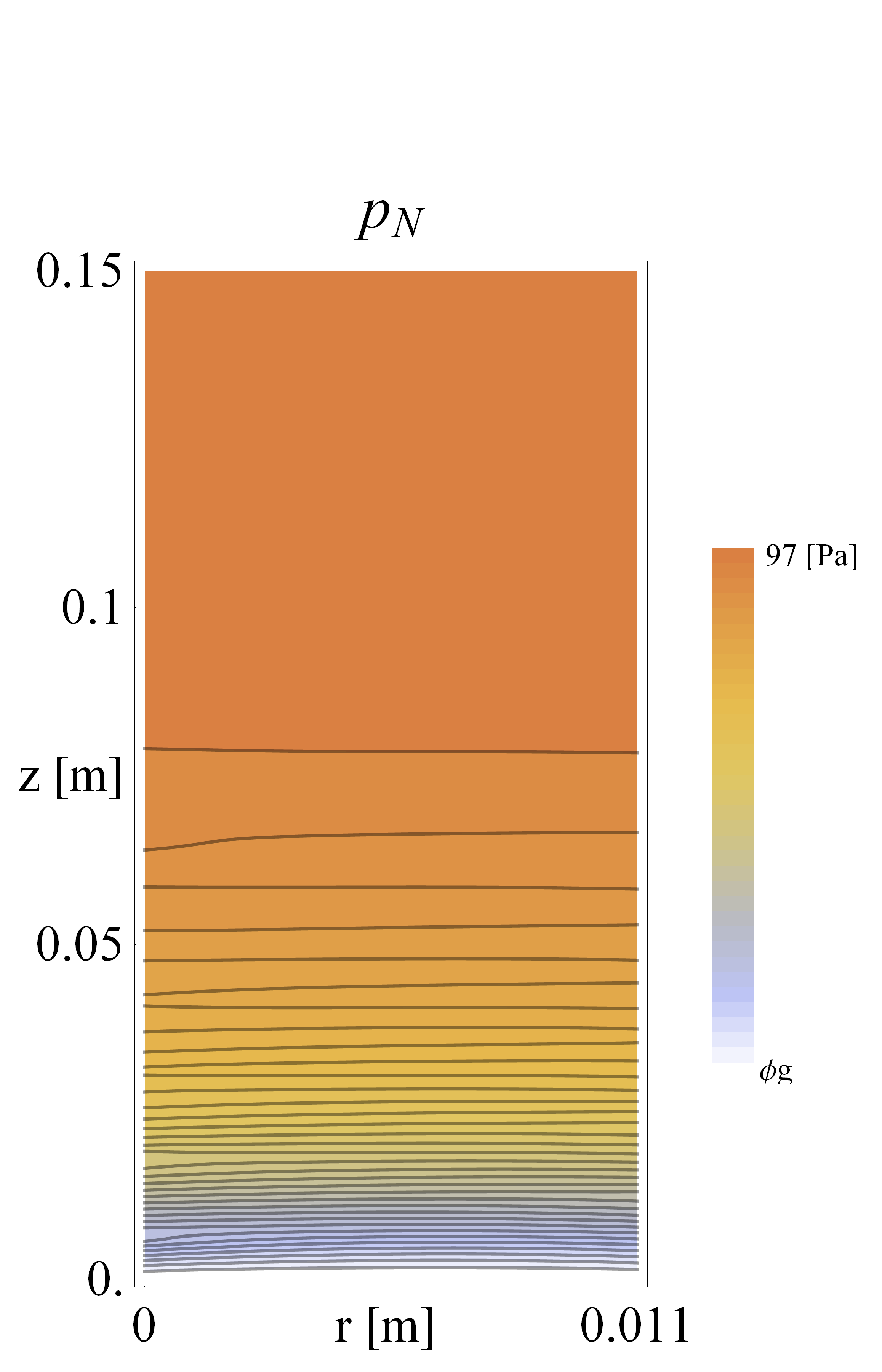}&
\includegraphics[width=0.32\columnwidth]{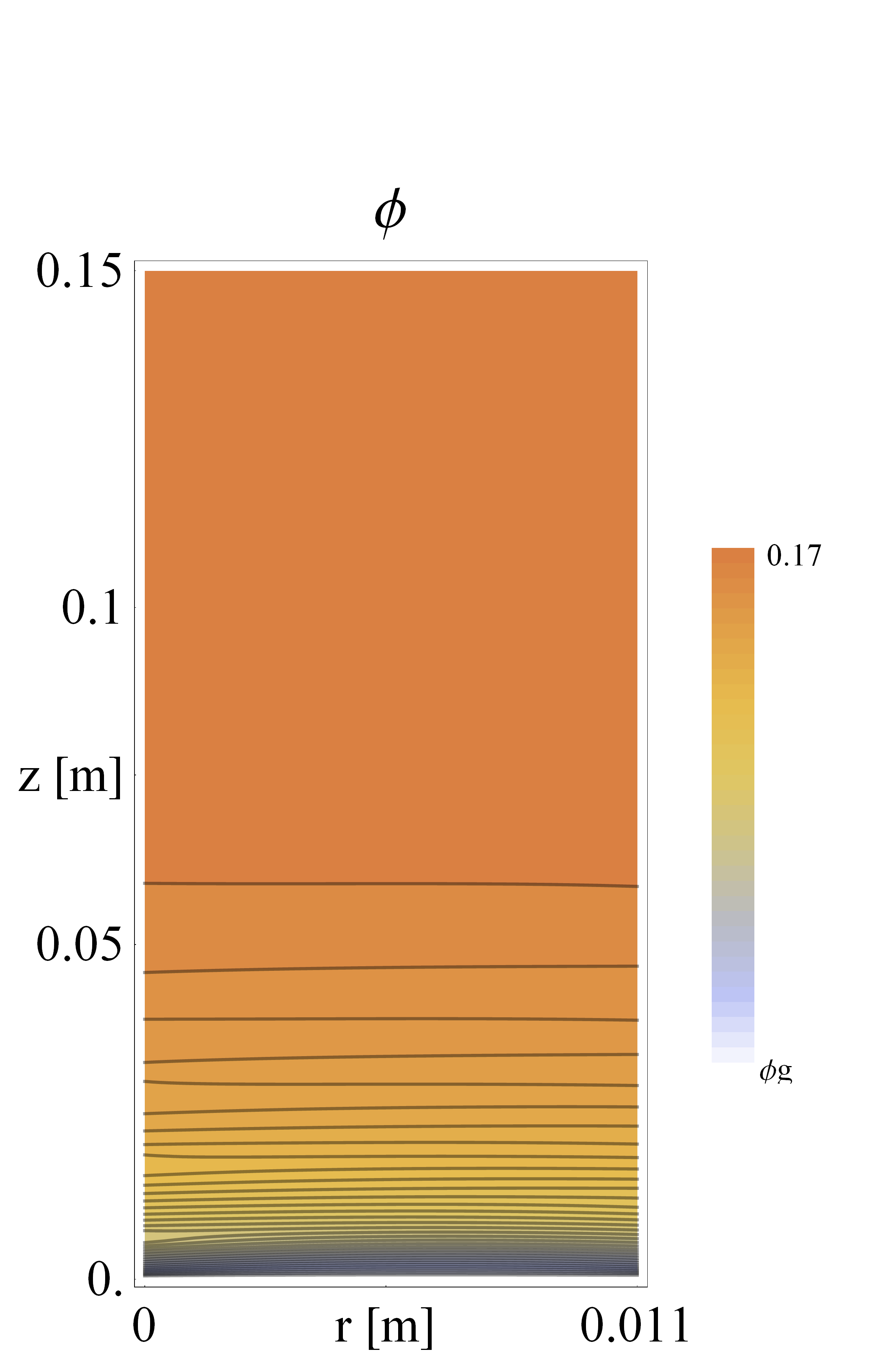}\\
(a) & (b) & (c)
\end{tabular}
\caption{Comparison of (a) network shear stress $\tau_N$, (b) network pressure $p_N$ and solids volume fraction $\phi_\infty$ distributions for hyper-elastic constitutive model.}\label{fig:hyperelastic}
\end{figure}

Conversely, the stress distribution for the hyperelastic solution (see \cite{LesterEA:14} for details) shown in Fig.~\ref{fig:hyperelastic}(a) has the same gross form as the visco-plastic solution, but the $C^1$ shocks are smoothed out, which appears to be a more physically realistic prediction. Whilst the shear stress distribution appears to be markedly different between the visco-plastic and hyper-elastic models, the corresponding pressure and solids volume fractions are remarkably similar (Fig.~\ref{fig:viscoplastic}(b)-(c), Fig.~\ref{fig:hyperelastic}(b)-(c)). As such, despite the $C^1$ shocks in the shear stress distribution, the visco-plastic constitutive model generates a highly accurate (within 1-2\% relative error) approximation to the equilibrium solids volume fraction distribution.

\section{Analytic 1D Approximation}\label{sec:1Dapprox}

Whilst the visco-plastic model provides a good approximation to the full hyper-elastic solution, and may be solved numerically in a straightforward manner, no analytic solution is known for the hyperbolic system (\ref{eqn:radial}), (\ref{eqn:vertical}), hence computationally intensive numerical iteration is still required to fit experimental data. As we seek a simplified analytic solution which can be employed routinely to analyse equilibrium batch sedimentation data, an accurate 1D approximation to the multi-dimensional problem is sought. To develop such an approximation, we consider the cross-sectional average
\begin{equation}
\bar{f}:=\frac{2}{R^2}\int_0^R r f(r) dr,
\end{equation}
of (\ref{eqn:vertical}), yielding the averaged force balance
\begin{equation}
\frac{d\bar{p}_N}{dz}+\frac{2}{R}\tau_y(\phi_\infty|_{r=R})-\Delta\rho g \bar\phi_\infty=0.\label{eqn:1Dexact}
\end{equation}
This force balance is very similar to (\ref{eqn:1Dbalance}) proposed by \citet{Michaels/Bolger:62}, but it differs in that (\ref{eqn:1Dexact}) is exact (up to the assumption that $\tau_y\approx\tau_w$), where in general $\bar\phi_\infty\neq\phi_\infty|_{r=R}$. We express the deviation of $\phi$ from its cross-sectional average $\bar\phi_\infty$ as
\begin{equation}
\phi_\infty=\bar\phi_\infty-\delta\phi_\infty,
\end{equation}
and determine the magnitude of $\delta\phi_\infty$ as follows. As the shear yield strength $\tau_y(\phi)$ is significantly smaller than the compressive yield stress $P_y(\phi)$ for volume fractions $\phi$ greater than the gel point, then typically $\tau_N\ll p_N$, and so from (\ref{eqn:vertical}) and boundary conditions (\ref{eqn:shearr0}), (\ref{eqn:shearrR}), $p_N$ only varies radially very weakly, as is shown in Fig.~\ref{fig:viscoplastic} (b). Furthermore, as $P_y(\phi)$ is a strongly increasing (roughly power-law) function of $\phi$, then $\phi_\infty$ varies even more weakly with radius, hence $\delta\phi_\infty$ is considered negligible, as is reflected in Fig.~\ref{fig:viscoplastic} (c). As such, the approximation
\begin{equation}
\phi_\infty|_{r=R}\approx\bar\phi_\infty\approx P_y^{-1}(\bar{p}_N),\label{eqn:deltaphi}
\end{equation}
is invoked in (\ref{eqn:1Dexact}), yielding an integral equation for the average network pressure as a function of bed depth $z$ in terms of the rheological functions $P_y(\phi)$, $\tau_y(\phi)$ as
\begin{equation}
z\approx\int_0^{\bar{p}_N}\frac{dp'}{\Delta\rho g P_y^{-1}(p')-\frac{2}{R}F_1(p')}.\label{eqn:integral}
\end{equation}

\begin{figure}[]
\centering
\includegraphics[width=0.4\columnwidth]{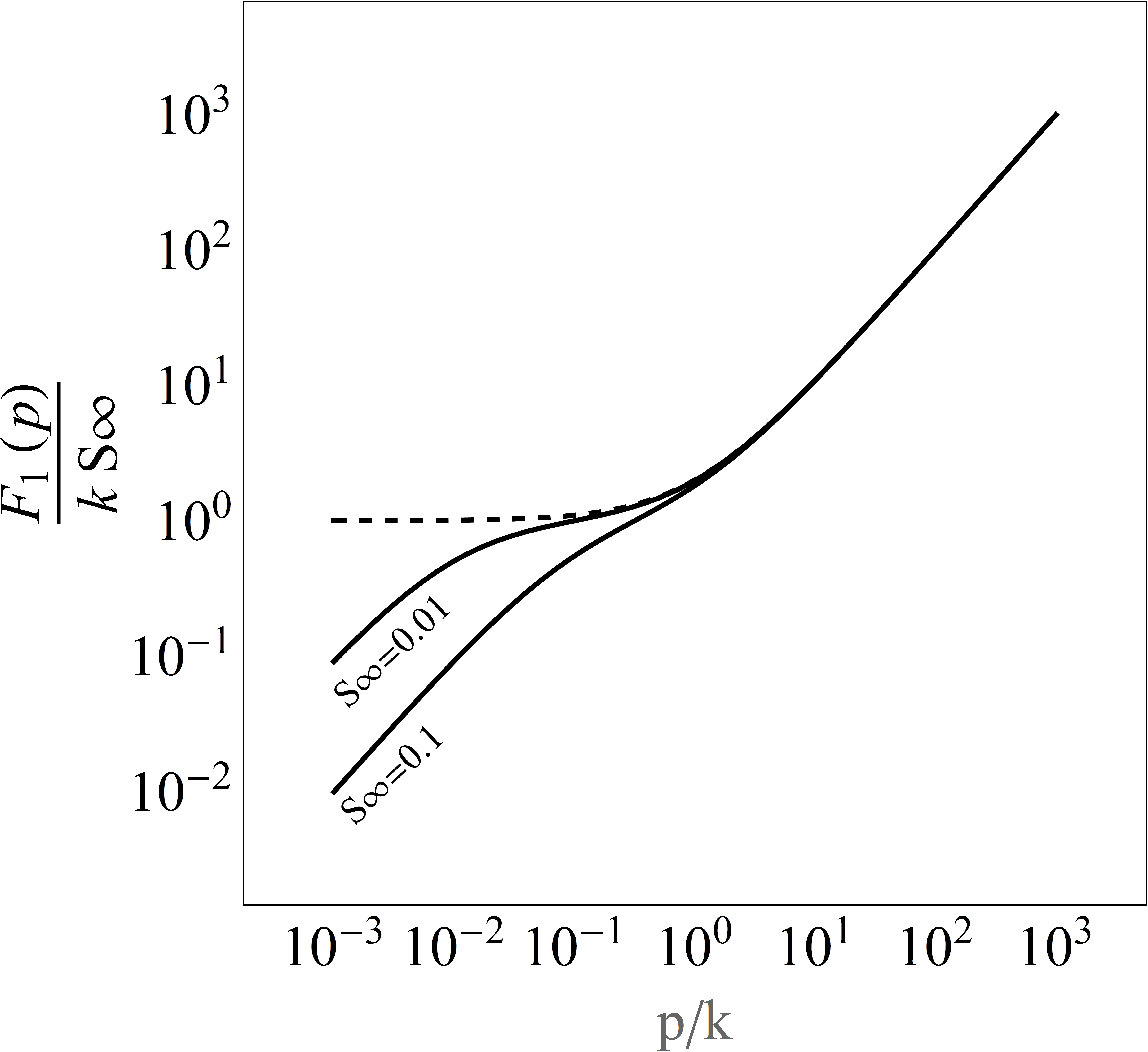}
\caption{Approximation of the function $F_1(p)$ (solid) for various values of $S_\infty$ with $S_\infty\left(\frac{p}{k}+1\right)$ (solid).}\label{fig:Fp_approx}
\end{figure}

Whilst this integral does not have an analytic solution for the functional forms of $P_y(\phi)$, $\tau_y(\phi)$ given by (\ref{eqn:pyphi_func}), (\ref{eqn:FpN}), an analytic solution is generated when $F_1(p)$ is approximated by the simple function
\begin{equation}
F_1(p)\approx S_\infty\left(\frac{p}{k}+1\right),\label{eqn:approxF}
\end{equation}
which, as shown in Fig.~\ref{fig:Fp_approx}, is a good approximation for $p/k>S_\infty$. Under this approximation, (\ref{eqn:integral}) is then
\begin{equation}
z\approx\int_0^{\bar{p}}\frac{dp'}{\Delta\rho g \phi_g\left(\frac{p'}{k}+1\right)^{\frac{1}{n}}-\frac{2 S_\infty}{R}\left(\frac{p'}{k}+1\right)},
\end{equation}
which may be integrated analytically to yield approximate expressions for both the average network pressure $\bar{p}_N$
\begin{equation}
\bar{p}_N(z)\approx k\left[\frac{\Delta\rho g\phi_g R}{2 S_\infty k}+\left(1-\frac{\Delta\rho g\phi_g R}{2 S_\infty k}\right)\exp\left(-\frac{n-1}{n}\frac{2 S_\infty}{R}z\right)\right]^{\frac{n}{n-1}}-k,\label{eqn:barp_approx}
\end{equation}
and equilibrium solids volume fraction $\bar\phi_\infty$
\begin{equation}
\bar{\phi}_\infty(z)\approx \phi_g\left[\frac{\Delta\rho g\phi_g R}{2 S_\infty k}+\left(1-\frac{\Delta\rho g\phi_g R}{2 S_\infty k}\right)\exp\left(-\frac{n-1}{n}\frac{2 S_\infty}{R}z\right)\right]^{\frac{1}{n-1}},\label{eqn:barphi_approx}
\end{equation}

\begin{figure}[]
\centering
\begin{tabular}{c c c}
\includegraphics[width=0.32\columnwidth]{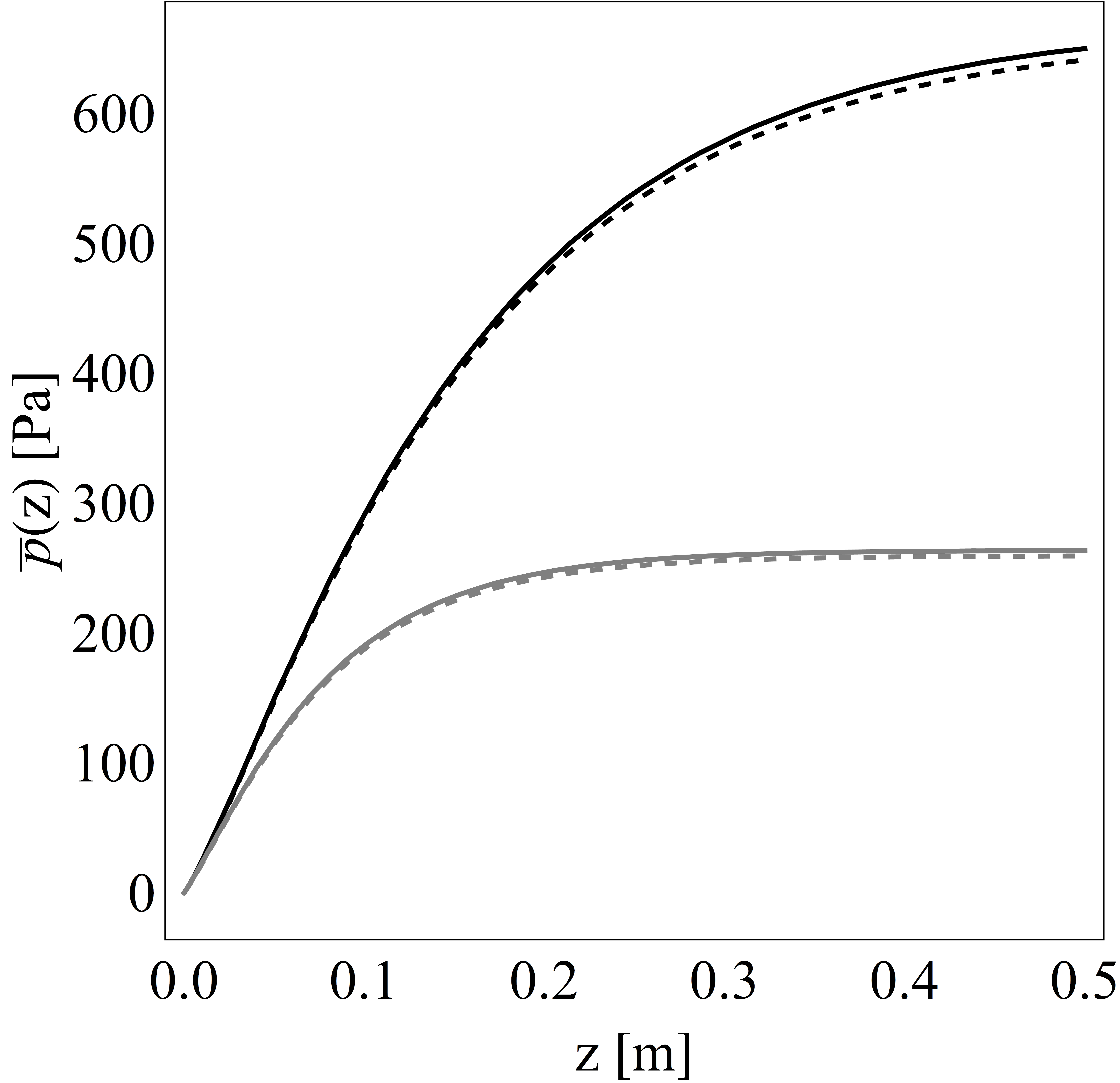}&
\includegraphics[width=0.32\columnwidth]{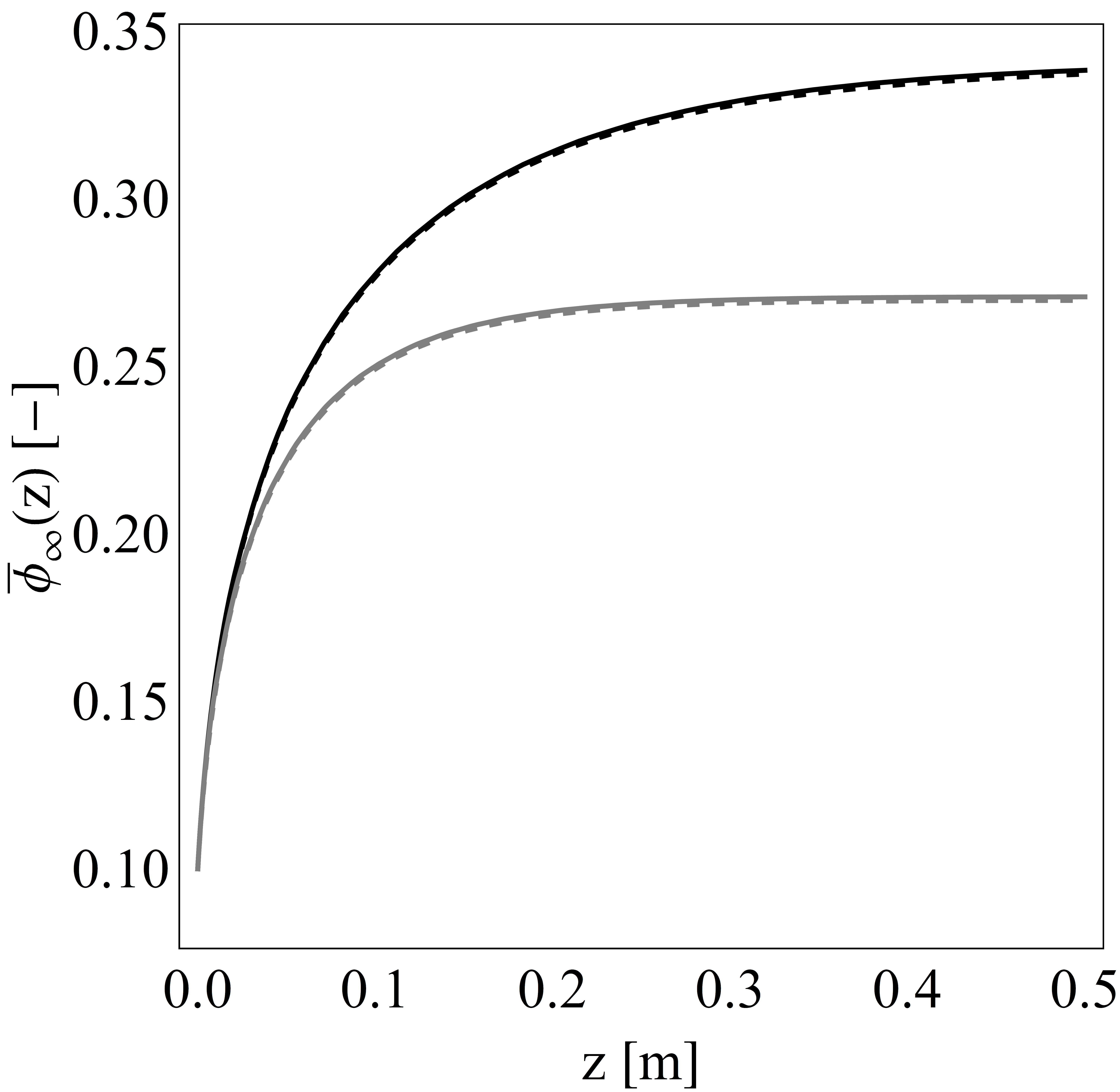}&
\includegraphics[width=0.32\columnwidth]{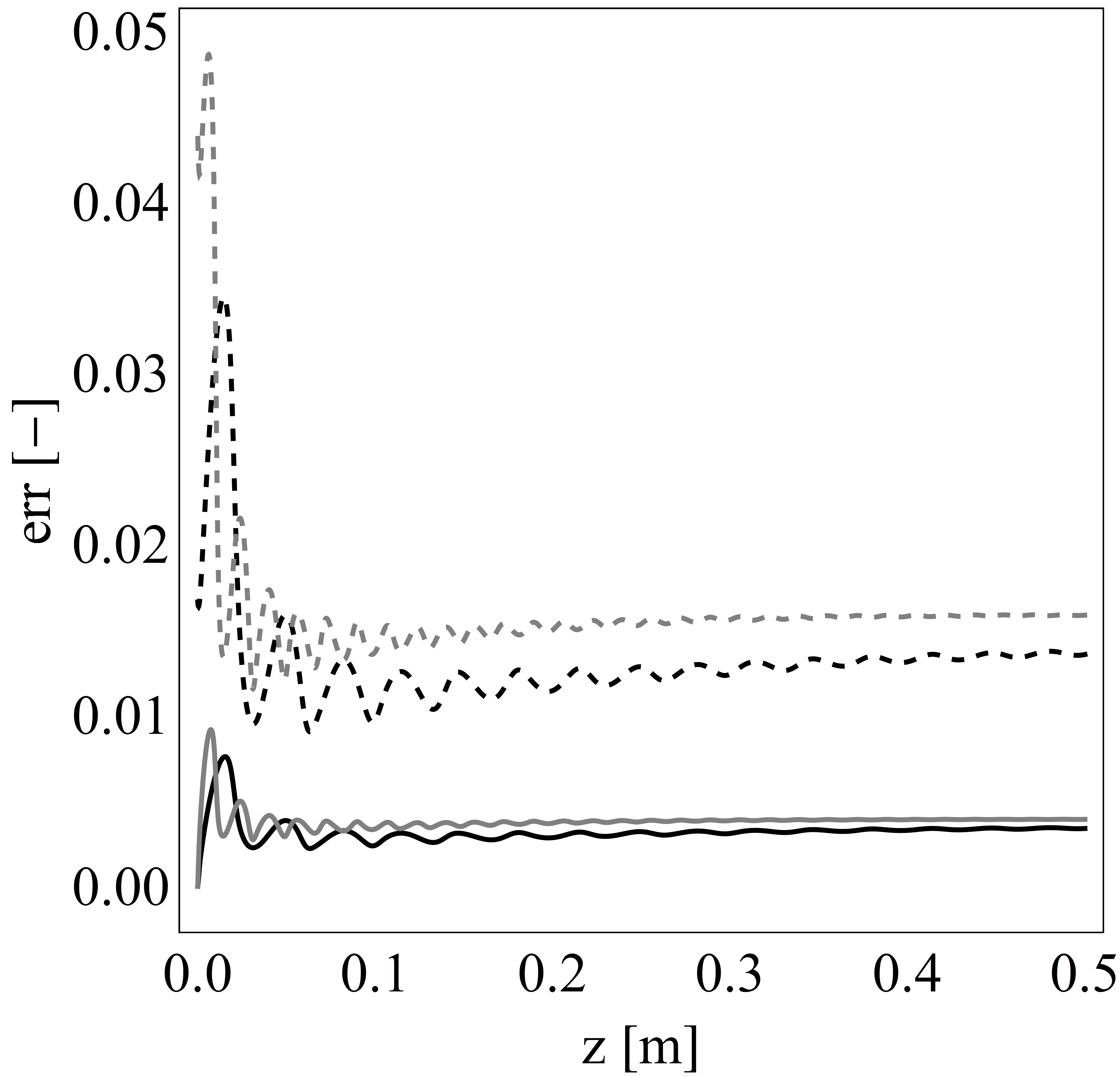}\\
(a) & (b) & (c)
\end{tabular}
\caption{Comparison of typical predictions of (a) average network pressure $\bar{p}_N$ and (b) average equilibrium solids volume fraction $\bar\phi_\infty(z)$ for $R_s$=0.011 [m] (gray) and $R_l$=0.055 [m] (black) column widths and (c) error summary for network pressure (dashed) and solids volume fraction (solid) between 2D visco-plastic and 1D approximate solutions of equilibrium stress state.}\label{fig:1Dcomparison}
\end{figure}

which satisfies $\lim_{z\rightarrow\infty}\bar{\phi}_\infty\rightarrow\phi_c$. As shown in Fig.~\ref{fig:1Dcomparison}, these expressions represent quite accurate approximations to both the 2D visco-plastic and 2D hyper-elastic solutions, with typical errors for $\bar\phi_\infty(z)$ and $\bar{p}_N(z)$ less than 1\% and 1-5\% respectively. These errors decay with increasing $z$ due to increasing accuracy of the approximations (\ref{eqn:deltaphi}) and (\ref{eqn:approxF}) with $p_N$. These closed-form 1D approximations (\ref{eqn:barp_approx}), (\ref{eqn:barphi_approx}) represent a significant simplification of both the multi-dimensional visco-plastic and hyper-elastic solutions which appear to be accurate enough for analysis of experimental data.

\section{Fitting of Equilibrium Solids Volume Fraction Profiles}\label{sec:volfrac}

Using (\ref{eqn:barp_approx}), (\ref{eqn:barphi_approx}), the inverse problem of estimation of the suspension rheology from a series of equilibrium solids volume fraction profiles is now greatly reduced to a simple nonlinear regression, where given such data and $\Delta\rho$, $g$, $R$, the rheological parameters $\phi_g$, $k$, $n$, $S_\infty$ which define the compressive and shear yield strengths can be directly estimated. To test the accuracy of this method, equation (\ref{eqn:barphi_approx}) is fitted to a series of equilibrium solids volume fraction profiles presented in \cite{LesterEA:14} for three different polymer-flocculated colloidal suspensions. These suspensions consist of 4 micron calcium carbonate primary particles under different flocculation conditions as summarized in Table~\ref{tab:suspensions} (see \cite{LesterEA:14} for further details). Equilibrium solids volume fraction profiles for each of these suspensions were measured using gamma ray attenuation~\cite{LabbettEA:06} in two different width columns; a narrow column $R_s$=0.011 [m], and a wide column $R_l$=0.055 [m].

\begin{figure}[h]
\centering
\begin{tabular}{c c c}
\includegraphics[width=0.32\columnwidth]{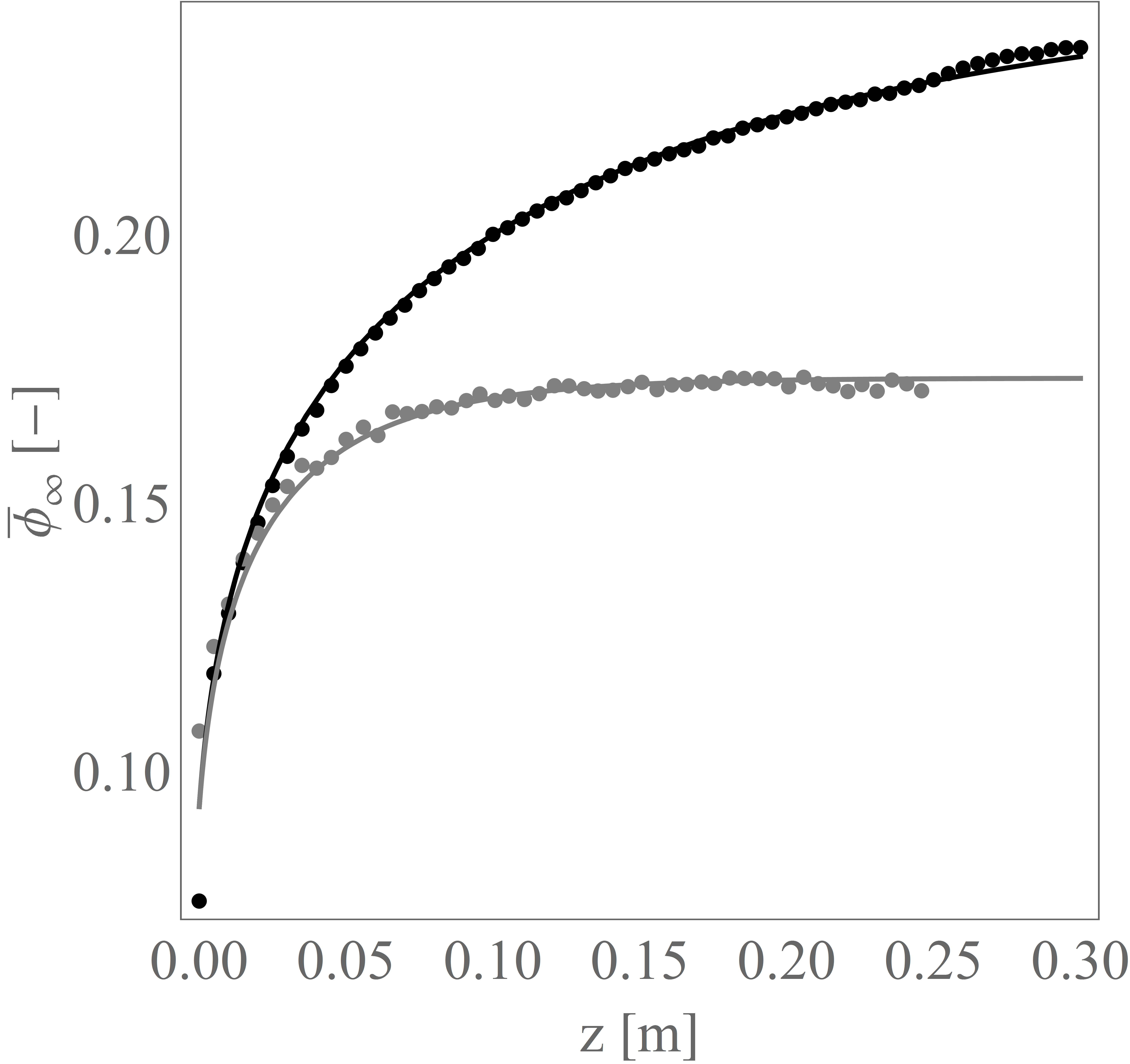}&
\includegraphics[width=0.32\columnwidth]{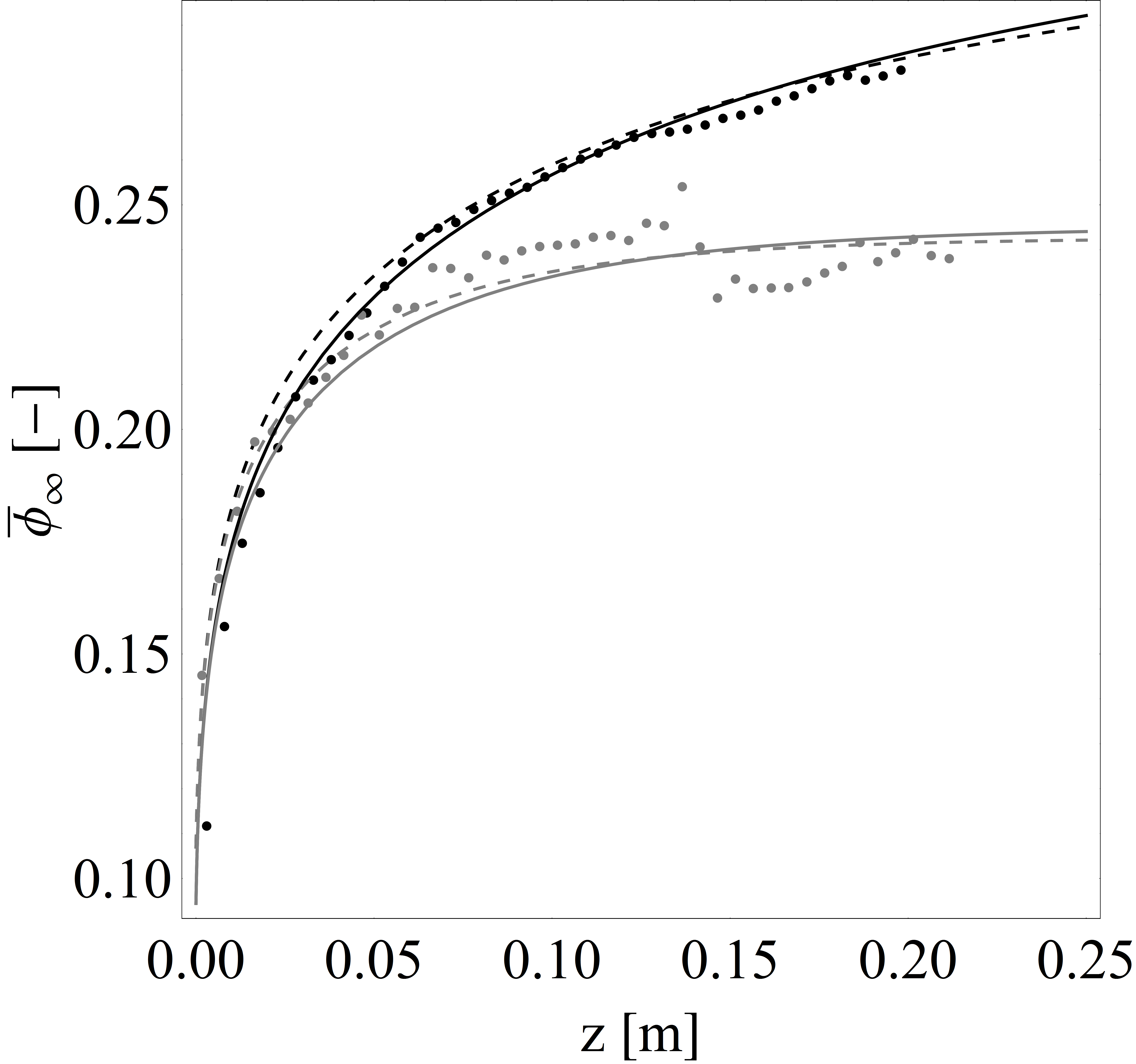}&
\includegraphics[width=0.32\columnwidth]{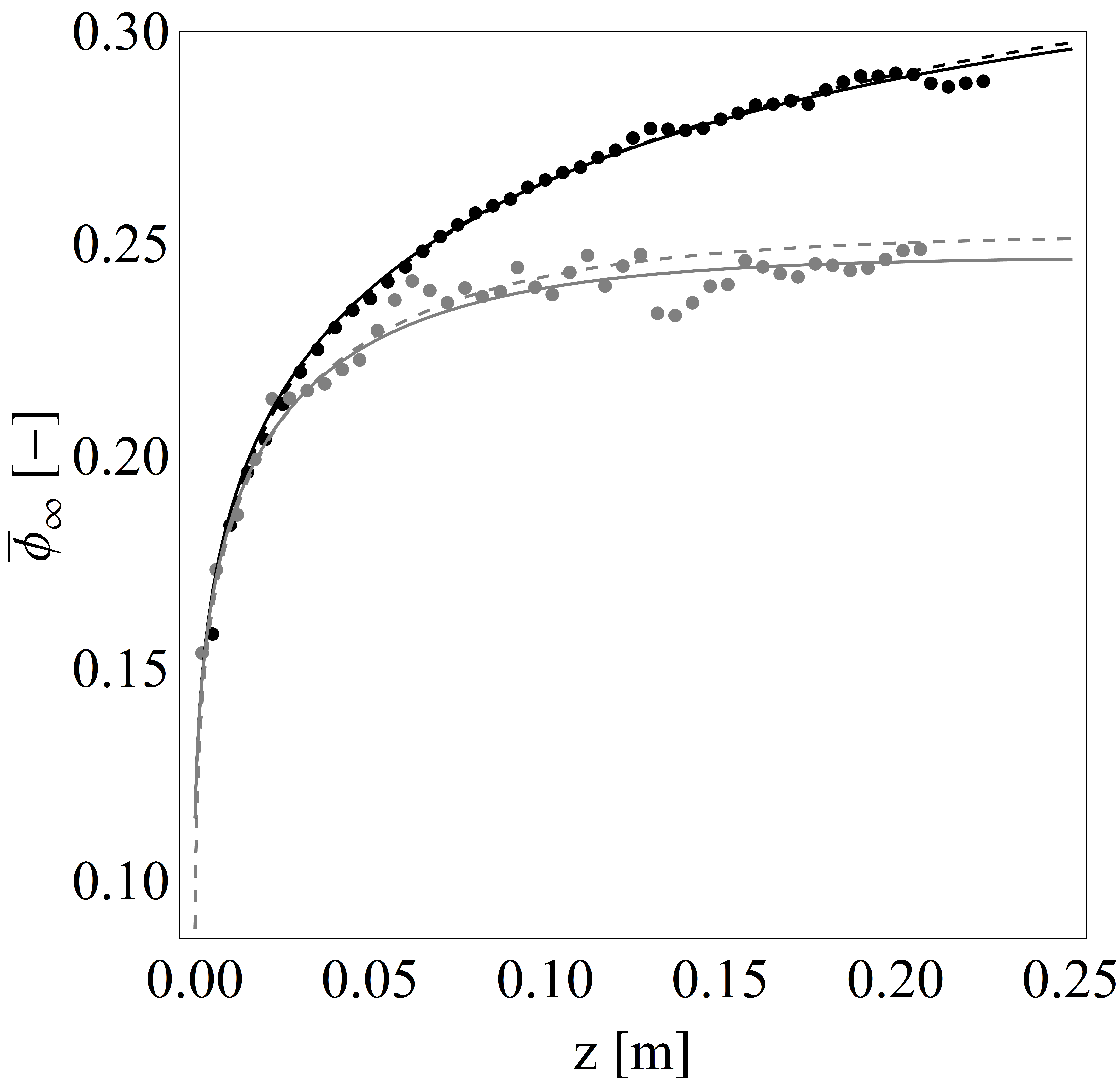}\\
(a) & (b) & (c)
\end{tabular}
\caption{Measured data (points) and model fits (1D model - solid lines, 2D model - dashed lines) of the equilibrium solids volume
fraction profile $\bar\phi_\infty(z)$ for $R_s$=0.011 [m] (gray) and $R_l$=0.055 [m] (black) column widths for calcium carbonate suspensions under flocculant types and dosages (a)-(c) summarized in Table~\ref{tab:suspensions}.}\label{fig:datafitting}
\end{figure}

\begin{table}[b]
\begin{centering}
\begin{tabular}{c c c c c c c}
\hline
Suspension & Method & $\phi_g$ [-] & $k$ [Pa] & $n$ [-] & $S_\infty$ [-] & $k/\phi_g^n$ [Pa] \\
\hline
(a) Magnafloc 46 [g/t] & 2D & 0.0918 & 3.211 & 5.482 & 0.1572 & 1.557$\times$10$^6$ \\
(a) Magnafloc 46 [g/t] & 1D & 0.0923 & 3.204 & 5.495 & 0.1597 & 1.556$\times$10$^6$ \\
(b) Rheomax 30 [g/t] & 2D & 0.1042 &  0.631 & 7.031 & 0.1121 & 4.211$\times$10$^6$ \\
(b) Rheomax 30 [g/t] & 1D & 0.1017 &  0.476 & 7.008 & 0.0966 & 4.308$\times$10$^6$ \\
(c) Rheomax 46 [g/t] & 2D & 0.0890 & 0.1627 & 7.014 & 0.1132 & 3.805$\times$10$^6$ \\
(c) Rheomax 46 [g/t] & 1D & 0.1150 & 0.9395 & 7.001 & 0.1159 & 3.540$\times$10$^6$ \\
\hline
\end{tabular}
\caption{Suspension flocculant type, dosage and fitted rheological parameters.}\label{tab:suspensions}
\end{centering}
\end{table}

\begin{figure}[h]
\centering
\begin{tabular}{c c c}
\includegraphics[width=0.32\columnwidth]{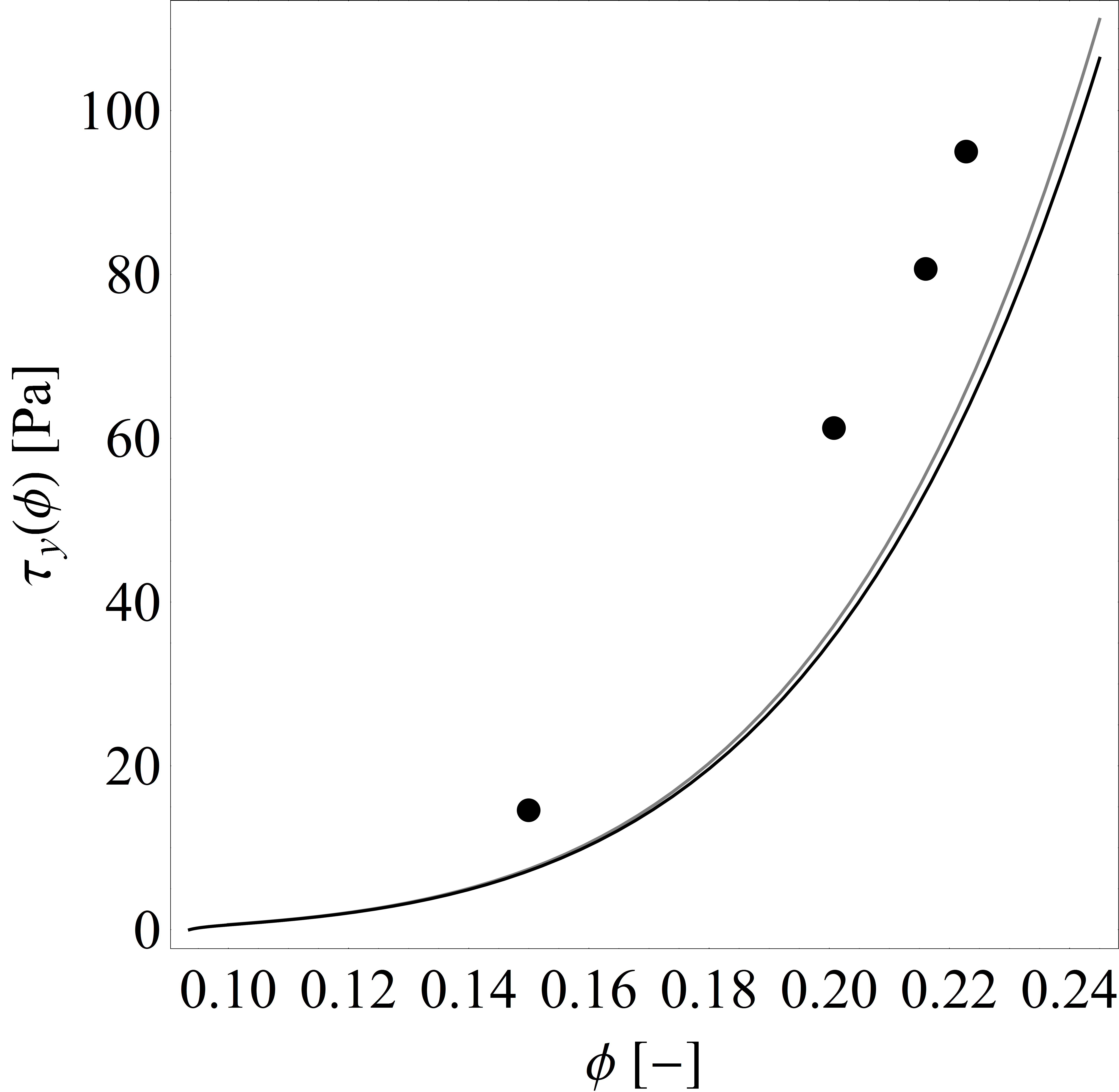}&
\includegraphics[width=0.32\columnwidth]{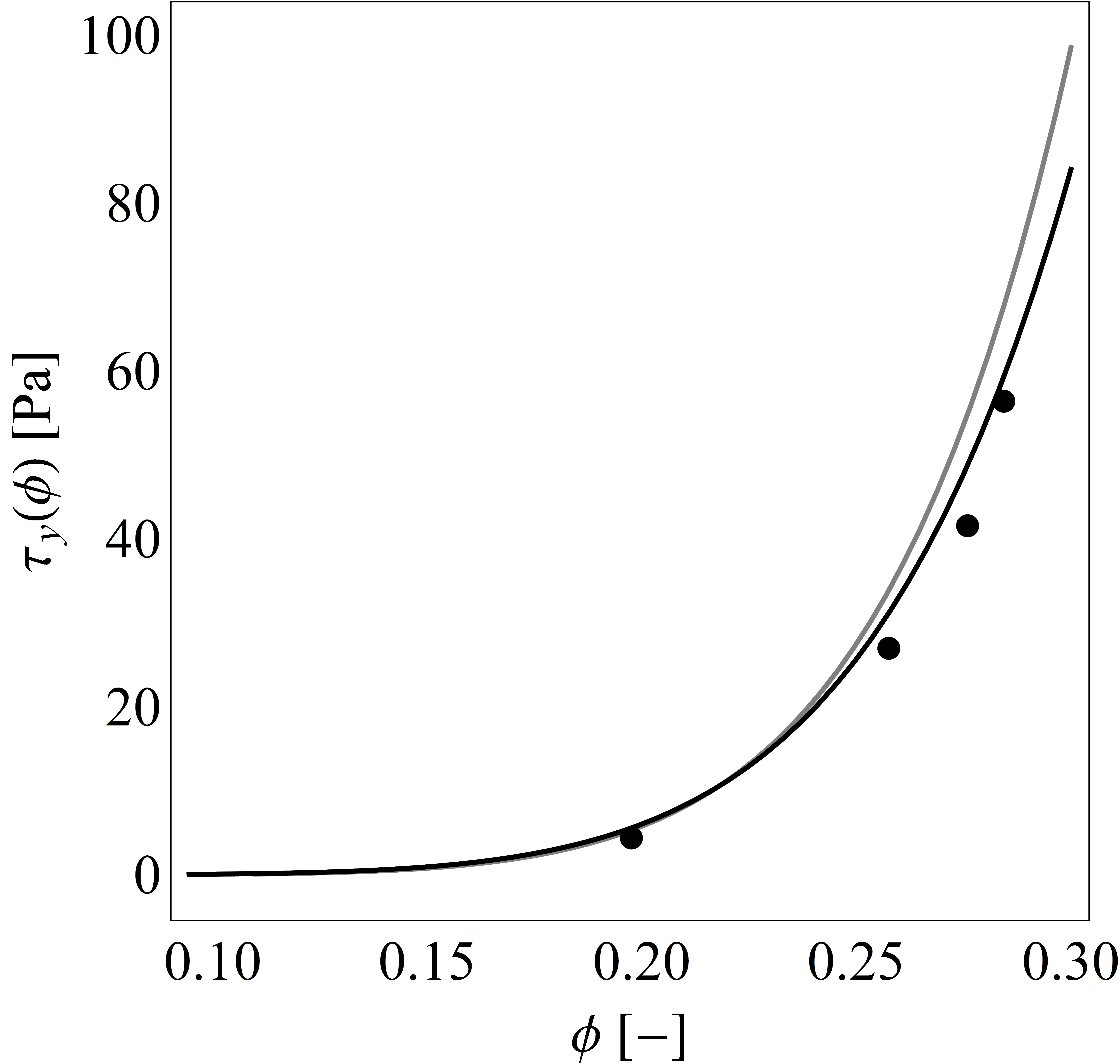}&
\includegraphics[width=0.32\columnwidth]{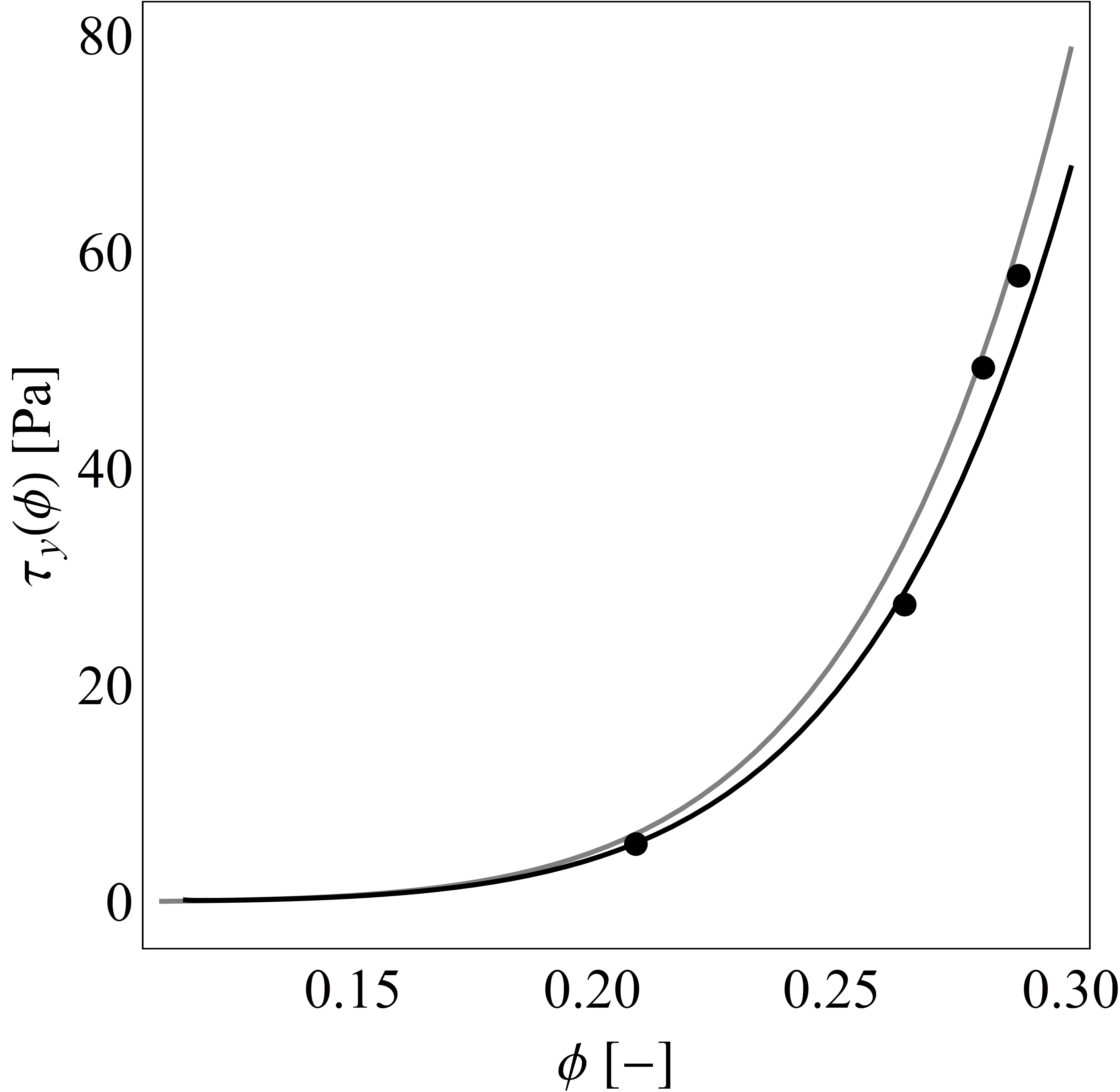}\\
(a) & (b) & (c)
\end{tabular}
\caption{Comparison between in-situ measurements (points) and model predictions (1D model - black lines, 2D model - gray lines) of the shear yield stress $\tau_y(\phi)$ for suspensions (a)-(c) summarized in Table~\ref{tab:suspensions}.}\label{fig:tau_est}
\end{figure}

The measured solids volume fraction profile data and fitted profiles using both the 1D analytic approximation (\ref{eqn:barphi_approx}) and 2D visco-plastic model are shown in Fig.~\ref{fig:datafitting}, which indicate both methods provide good fits to the experimental data, as reflected by the fitted rheological parameters shown in Table~\ref{tab:suspensions}. The agreement between the fitted parameters is excellent for the Magnafloc suspension, whereas it is only reasonable for the two Rheomax suspensions. As shown in Table~\ref{tab:suspensions}, these poorer fits are attributed to ambiguity in deconvolution of $k$ and $\phi_g$ in the term $k/\phi_g^n$ from the functional form of $P_y(\phi)$ in (\ref{eqn:pyphi_func}). This behaviour is typical of estimation of the compressive yield stress in general, as the suspension gel point $\phi_g$ represents an asymptotic limit of vanishing stress, and hence it is difficult to predict from equilibrium solids profiles alone~\cite{Green/Landman:96}. Hence, whilst the 1D and 2D methods can generate different estimates of the individual rheological parameters, the estimated rheological functions $P_y(\phi)$, $\tau_y(\phi)$ are quantitatively very similar, as shown in Fig.s~\ref{fig:datafitting} and \ref{fig:tau_est}.

The fitted rheological parameters can also be used to compare the estimated shear yield strength (under the assumption that this is similar to the wall adhesion strength) with \emph{in-situ} measurements as performed in \cite{LesterEA:14}. A comparison between the measured data and predictions from the 1D analytic and 2D numerical visco-plastic models is shown in Fig.~\ref{fig:tau_est}. In this case, the agreement between the 1D and 2D fitted curves for $\tau_y(\phi)$ is very good for all suspensions tested. These results suggest that the 1D approximation (\ref{eqn:barphi_approx}) is capable of generating accurate estimates of the both the compressive and shear yield strength of colloidal suspensions from equilibrium solids volume fraction profile data. Note that application of the 1D or 2D methods to either the narrow or wide column solids volume fraction profile data alone results in unacceptably large errors in the estimated rheological parameters, reinforcing the notion that several columns of different width are required to deconvolute contributions from compression and wall adhesion.


\begin{figure}[h]
\centering
\begin{tabular}{c c}
\includegraphics[width=0.48\columnwidth]{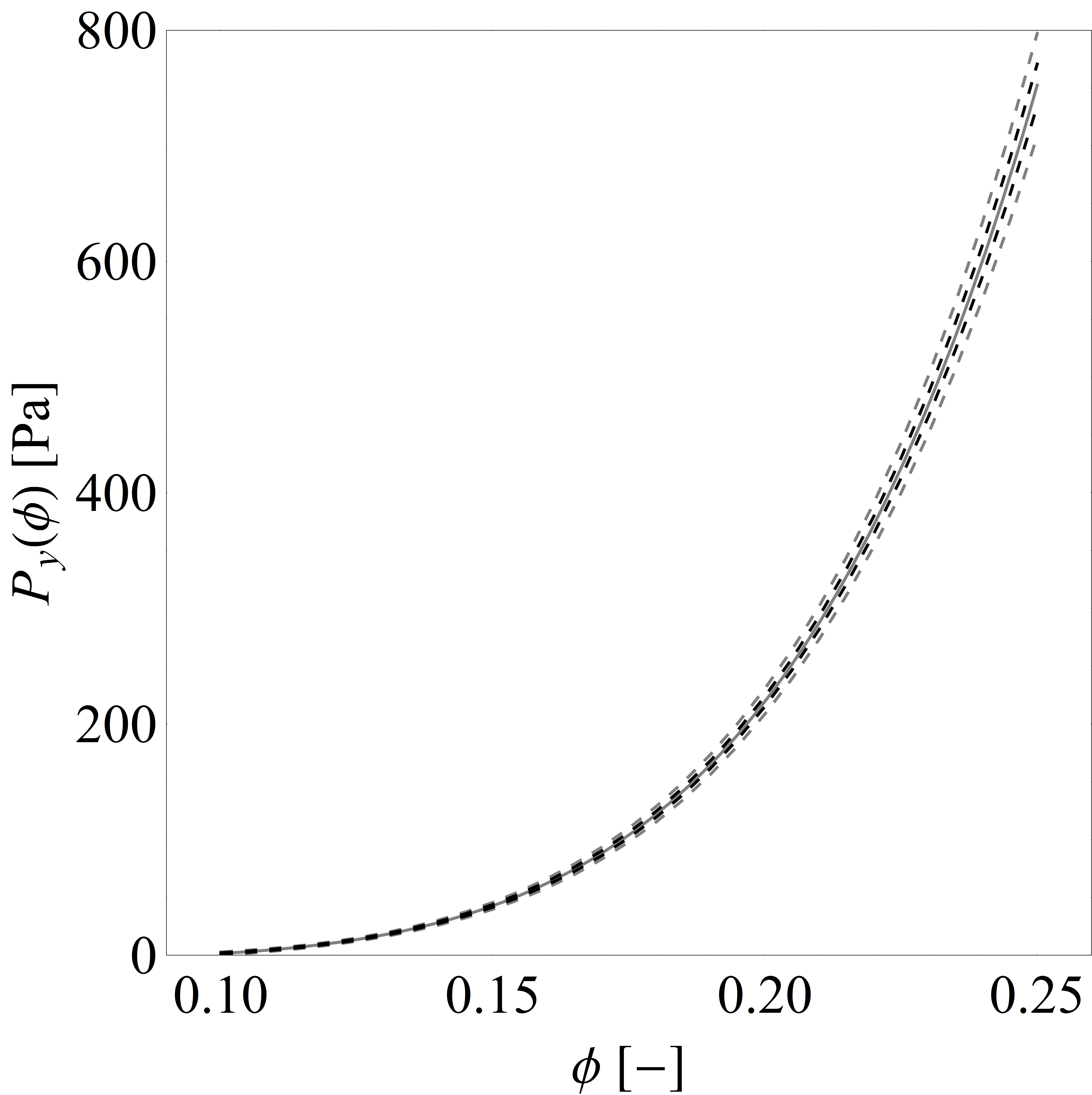}&
\includegraphics[width=0.48\columnwidth]{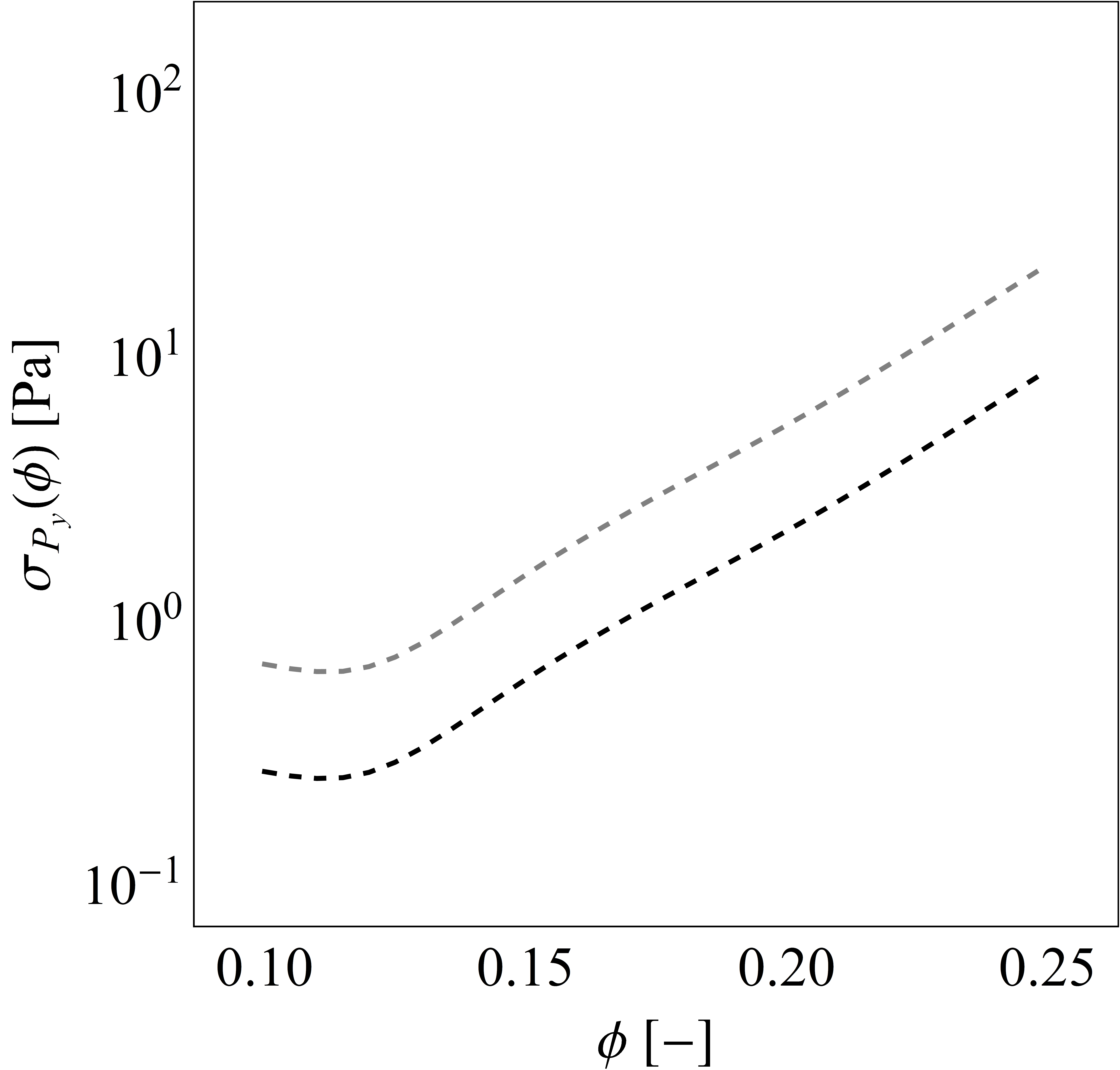}\\
(a) & (b)
\end{tabular}
\caption{(a) Compressive yield strength $P_y(\phi)$ curve for suspension (a) (solid, gray), and 95\% confidence intervals for $\sigma_{\bar\phi_\infty}$=0.002 (dashed, black) and $\sigma_{\bar\phi_\infty}$=0.005 (dashed, gray), and (b) plots of standard deviation $\sigma_{P_y}(\phi)$ of the compressive yield strength for $\sigma_{\bar\phi_\infty}$=0.002 (dashed, black) and $\sigma_{\bar\phi_\infty}$=0.005 (dashed, gray).}\label{fig:Py_errors}
\end{figure}

To gauge the sensitivity of the 1D regression method to measurement errors, Gaussian noise with zero mean and standard deviation $\sigma_{\bar\phi_\infty}$ is added to the experimental solids volume fraction profile data for the Magnafloc suspension shown in Fig.~\ref{fig:datafitting}(a). Two different levels of Gaussian noise $\sigma_{\bar\phi_\infty}$=0.002, $\sigma_{\bar\phi_\infty}$=0.005 are used to generate $10^4$ random realizations of experimental data, and the resultant mean $\mu_\epsilon$ and standard deviation $\sigma_\epsilon$ of the relative error of the fitted rheological parameters $\phi_g$, $k$, $n$, $S_\infty$ are summarized in Table~\ref{tab:sensitivity}. In both cases the mean $\mu_\epsilon$ is small for all of the rheological parameters, and the standard deviation $\sigma_\epsilon$ is small for all parameters except for the coefficient $k$. Again, this error appears to be related to the difficulty in deconvoluting the group $k/\phi_g^n$, as the relative error for this group (Table~\ref{tab:sensitivity}) is universally small. This is also reflected in Fig.~\ref{fig:Py_errors}, where the range of estimates of the compressive yield stress are excellent for both values of $\sigma_{\bar\phi_\infty}$, and are much smaller than suggested by the errors associated with the parameters $k$, $\phi_g$ alone.

\begin{table}[b]
\begin{centering}
\begin{tabular}{c c c c c c}
\hline
 & $\sigma_{\bar{\phi}_\infty}$ & $\phi_g$ & $S_\infty$ & $k$ & $n$ \\
\hline
$\mu_\epsilon$ & 0.005 & -0.032\% & -0.008\% & 2.564\% & -0.008\% \\
$\sigma_\epsilon$ & 0.005 & 3.681\% & 4.727\% & 25.56\% & 4.728\% \\
$\mu_\epsilon$ & 0.002 & -0.015\% & 0.004\% & 0.356\% & 0.004\% \\
$\sigma_\epsilon$ & 0.002 & 1.483\% & 1.894\% & 10.05\% & 1.893\% \\
\hline
\end{tabular}
\caption{Mean and standard deviation of relative error $\epsilon$ in estimates of rheological parameters $\phi_g$, $k$, $n$, $S_\infty$, based upon Magnafloc data for different values of solids volume fraction error $\sigma_{\bar\phi_\infty}$.}\label{tab:sensitivity}
\end{centering}
\end{table}

\section{Fitting of Equilibrium Bed Height Data}\label{sec:eqmheight}

In many instances the averaged solids volume fraction profile $\bar{\phi}_\infty(z)$ is not available experimentally, however it is simple to measure the equilibrium height $h_\infty$ over a range of experiments of different initial suspension height $h_0$ and column radius $R$. Estimation of the compressive yield strength $P_y(\phi)$ from equilibrium height data alone is not new \cite{Green/Landman:96}, and in the absence of wall adhesion effects, this procedure is relatively straightforward. In this section we derive relationships for the equilibrium height $h_\infty$ as a function of the column radius $R$ and the linear solids volume
\begin{equation}
m:=\phi_0 h_0=\int_0^{h_\infty} \bar{\phi}_\infty(z) dz,
\end{equation}
which facilitate estimation of the rheological parameters $\phi_g$, $k$, $n$, $S_\infty$ from a set of equilibrium height data. To derive this relationship, we directly integrate (\ref{eqn:barphi_approx}) to yield
\begin{equation}
\begin{split}
m(h_\infty)=& \frac{\phi_g(n-1)}{(q-1)r}\Big[-\,_2F_1\left(1,1,\frac{n-2}{n-1},\frac{q}{q-1}\right)\\
&+(1+q(e^{h_\infty r}-1))
(q+(1-q)e^{h_\infty r})^\frac{1}{n-1}
\,_2F_1\left(1,1,\frac{n-2}{n-1},\frac{q e^{h_\infty r}}{q-1}\right)\Big],\\
&\text{where}\,\,\,\,r=\frac{n-1}{n}\frac{2S_\infty}{R},\,\,q=\frac{\Delta\rho q\phi_g R}{2S_\infty k},\label{eqn:m_h_infty}
\end{split}
\end{equation}
and $\,_2F_1$ is the hypergeometric series
\begin{equation}
\,_2F_1(a,b;c;z):=\sum_{n=0}^\infty \frac{(a)_n(b)_n}{(c)_n}\frac{z^n}{n!},
\end{equation}
where $(x)_n$ is the Pochhammer symbol
\begin{equation}
(x)_n=
\begin{cases}
1\quad &n=0,\\
x(x+1)\cdots(x+n-1)\quad &n>0.
\end{cases}
\end{equation}

To test the accuracy of this method, we generated synthetic equilibrium height data for all combinations of 5 linear solids volumes $m=\phi_0h_0$=0.02-0.1 [m] in steps of 0.02 [m], and across 3 column radii $R$=(0.02,0.05,0.1)[m] based upon the rheological parameters derived for Magnafloc in Table~\ref{tab:suspensions}. A significant number (15) of experiments are required to generate a dense enough data set to provide reliable estimates of these parameters, and significantly wide columns are also required to deconvolute compression and wall adhesion effects. Note that for very dilute suspensions, the column heights required to achieve $m=\phi_0 h_0\sim 0.1 [m]$ may be very large, however as we are only interesting in equilibrium data, this problem may be alleviated by continually topping up settled beds and decanting supernatant until the requisite solids mass has been added, significantly reducing the maximum required column height.

\begin{table}[]
\begin{centering}
\begin{tabular}{c c c c c c c}
\hline
 & $\sigma_{h_\infty}$ & $\phi_g$ & $S_\infty$ & $k$ & $n$ & $k/\phi_g^n$ \\
\hline
$\mu_\epsilon$ & 0.5 [mm] & -3.665\% & 0.041\% & 10.60\% & -0.001\% & 0.247\%\\
$\sigma_\epsilon$ & 0.5 [mm] & 17.48\% & 4.041\% & 74.21\% & 1.044\% & 6.304\% \\
$\mu_\epsilon$ & 0.2 [mm] & -0.079\% & 0-0.079\% & 3.352\% & -0.0193\% & -0.051\% \\
$\sigma_\epsilon$ & 0.2 [mm] & 4.964\% & 1636\% & 29.33\% & 0.422\% & 2.528\% \\
\hline
\end{tabular}
\caption{Mean and standard deviation of relative error $\epsilon$ in estimates of rheological parameters $\phi_g$, $k$, $n$, $S_\infty$, based upon Magnafloc data for different values of height error $\sigma_{h_\infty}$.}\label{tab:error1}
\end{centering}
\end{table}

\begin{figure}[h]
\centering
\begin{tabular}{c c}
\includegraphics[width=0.48\columnwidth]{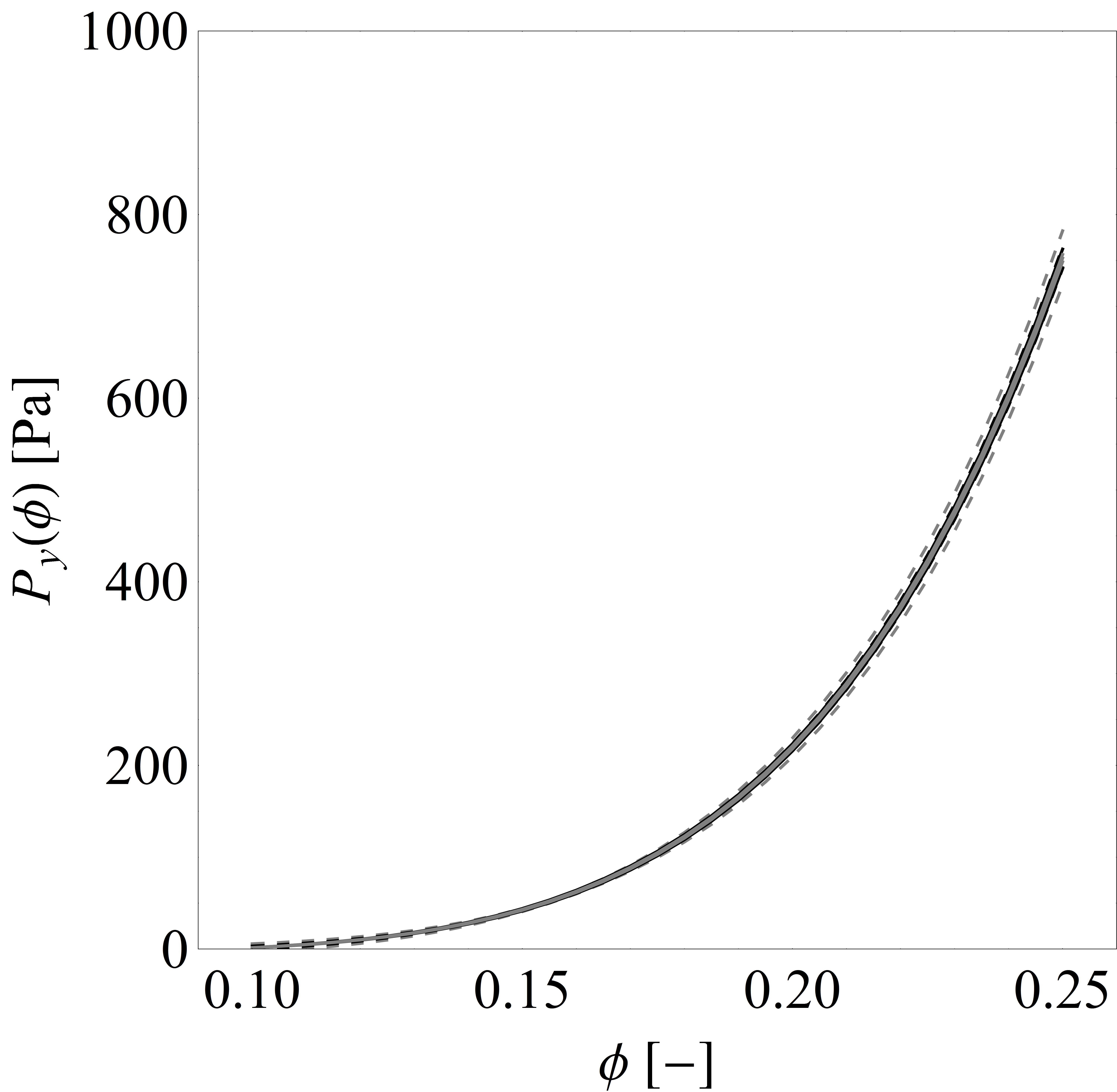}&
\includegraphics[width=0.48\columnwidth]{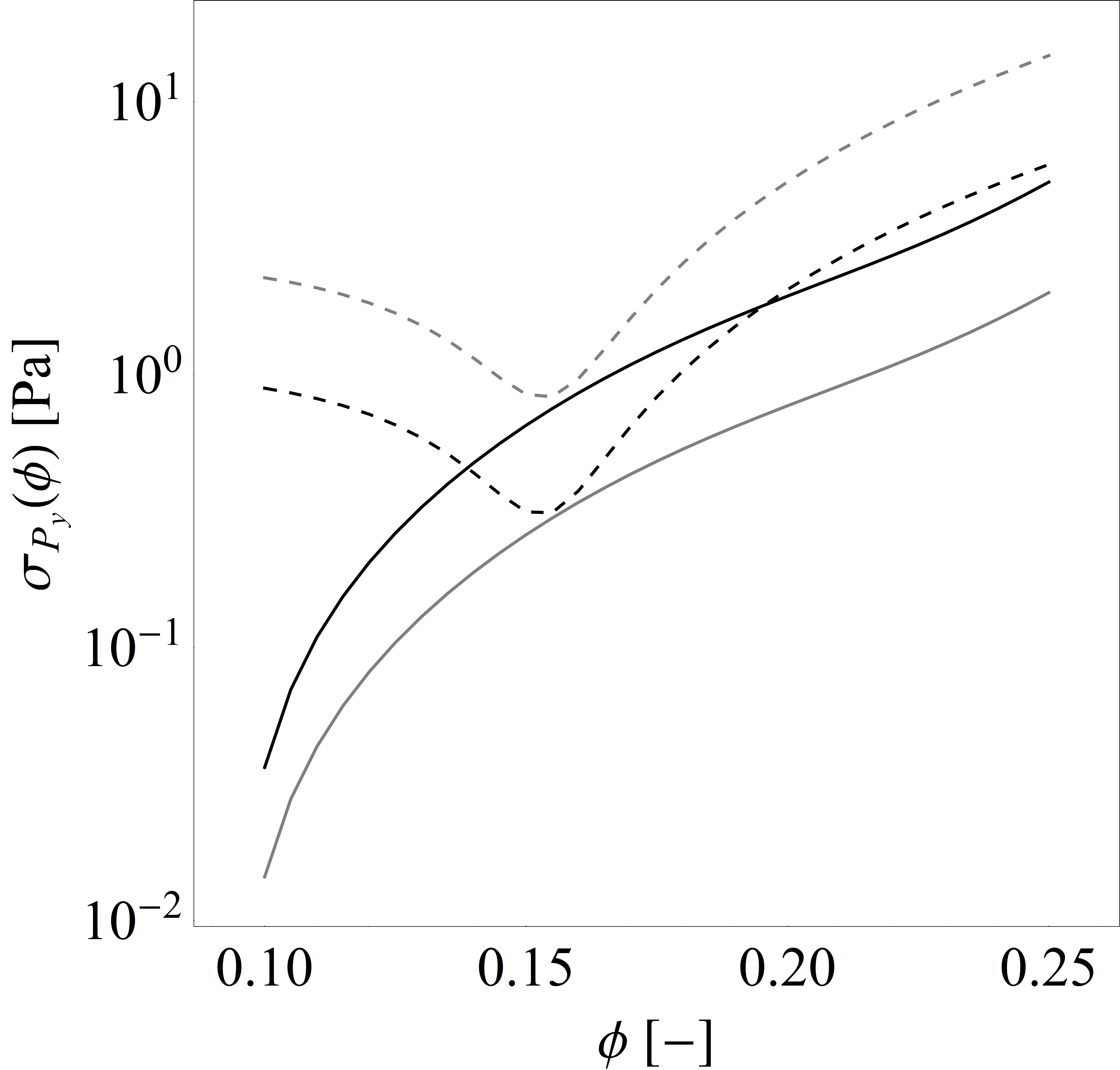}\\
(a) & (b)
\end{tabular}
\caption{(a) Compressive yield strength $P_y(\phi)$ curve for suspension (a) (solid, gray), and 95\% confidence intervals for $\sigma_{h_\infty}$=0.2 [mm] (dashed, black) and $\sigma_{h_\infty}$=0.5 [mm] (dashed, gray) when $\phi_g$ is unknown, and (b) plots of standard deviation $\sigma_{P_y}(\phi)$ of the compressive yield strength for $\sigma_{h_\infty}$=0.2 [mm] and $\phi_g$ unknown (dashed, black), $\sigma_{h_\infty}$=0.5 [mm] and $\phi_g$ unknown (dashed, gray), $\sigma_{h_\infty}$=0.2 [mm] and $\phi_g$ known (solid, black), $\sigma_{h_\infty}$=0.5 [mm] and $\phi_g$ known (solid, gray).}\label{fig:Py_errors2}
\end{figure}

To estimate the impact of errors in measuring the equilibrium height $h_\infty$, random Gaussian noise with zero mean and standard deviation $\sigma_{h_\infty}$ was added to the generated height data over $10^{4}$ samples, and the regression estimates recorded. The mean $\mu_\epsilon$ and standard deviation $\sigma_\epsilon$ of the relative error $\epsilon$ over this ensemble is shown in Table~\ref{tab:error1} for two noise levels $\sigma_{h_\infty}$=0.5[mm], $\sigma_{h_\infty}$=0.2[mm], and for the case $\sigma_{h_\infty}=0$ the rheological parameters in Table~\ref{tab:suspensions} are recovered to within machine precision. Whilst there appears to be only moderate skewness in the parameter estimates, as reflected by the mean $\mu_\epsilon$, the spread of errors as quantified by $\sigma_\epsilon$ is large, especially for the suspension gel point $\phi_g$ and compressibility parameter $k$. Again, this is reflected in the errors for the grouping $k/\phi_g^n$, which are significantly smaller that those of $\phi_g$ or $k$ alone, and indicate accurate estimates of the compressive yield strength can be made (Fig.~\ref{fig:Py_errors2}) although the individual parameter estimates may contain significant errors. In all cases, it is critical to make accurate equilibrium height measurements, as the difference between $\sigma_{h_\infty}=0.5 [mm]$ and $\sigma_{h_\infty}=0.2 [mm]$ is quite significant.

\begin{table}[b]
\begin{centering}
\begin{tabular}{c c c c c}
\hline
& $\sigma_{h_\infty}$ & $S_\infty$ & $k$ & $n$ \\
\hline
$\mu_\epsilon$ & 0.5 [mm] & 0.057\% & 0.027\% & -0.005\% \\
$\sigma_\epsilon$ & 0.5 [mm] & 1.542\% & 2.995\% & 0.555\% \\
$\mu_\epsilon$ & 0.2 [mm] & 0.044\% & -0.078\% & 0.015\% \\
$\sigma_\epsilon$ & 0.2 [mm] & 0.608\% & 1.189\% & 0.220\% \\
\hline
\end{tabular}
\caption{Mean and standard deviation of relative error $\epsilon$ in estimates of rheological parameters $k$, $n$, $S_\infty$, based upon Magnafloc data for different values of height error $\sigma_{h_\infty}$, assuming gel point $\phi_g$ is known.}\label{tab:error2}
\end{centering}
\end{table}

Often the suspension gel point $\phi_g$ has been obtained by other means, which constrains the parameter estimation problem. Such information resolves the deconvolution problem between $\phi_g$ and $k$, and significantly improves the accuracy of all parameter estimates. Whilst in principle the gel point can be estimated from equilibrium height data, estimates of $\phi_g$ from the column data given tend to over-estimate the gel-point, rather smaller values of $m$ are required to extrapolate toward the zero-pressure case.
The estimates for when $\phi_g$ is known are shown in Table~\ref{tab:error2}, and the relative errors are significantly reduced, less than 2.5\% for the 0.5[mm] error and 1\% for the 0.2[mm] error levels. These results in Fig.~\ref{fig:Py_errors2} suggest that it is possible to quite accurately estimate the compressive yield strength from equilibrium height data given careful experimentation and prior knowledge of the suspension gel point.

\section{Conclusions}\label{sec:conclusions}

Wall adhesion effects in equilibrium batch settling tests have been shown to be more prevalent than previously thought for strongly flocculated colloidal gels, and such artifacts can lead to unbounded errors in estimation of the compressive rheology of these materials. Indeed, wall effects are always likely to be important in gravity batch settling for two reasons, firstly, the effect of shear relative to compressive strength is amplified by a geometric factor of $2h_\infty/R$, and second, the suspension is usually not far from the gel-point where the ratio $S(\phi)$ of shear (adhesive) to compressive strength is unity, regardless of how small its asymptotic value $S_\infty$ is at high volume-fraction. In this study, we develop and test a 1D approximate solution to the equilibrium stress state which represents a significant simplification of previous multi-dimensional solution methods using either visco-plastic or hyper-elastic constitutive models. The 1D visco-plastic analytic solution provides accurate estimates (within several percent) of both the 2D visco-plastic and hyper-elastic solutions, and so is suitable for direct estimation of both the compressive $P_y(\phi)$ and shear $\tau_y(\phi)$ yield strength from equilibrium batch settling data via nonlinear regression.

Methods have been developed to estimate the shear and compressive rheology from either several equilibrium solids volume fraction profiles in different width settling columns, or a series of equilibrium height measurements over a combination of total solids and column radii. For equilibrium solids concentration profile data, accurate estimates of both $P_y(\phi)$ and $\tau_y(\phi)$ are possible, given the local cross-sectionally average equilibrium solids volume fraction $\bar\phi_\infty$ can be measured accurately, however there does exist an outstanding issue with respect to deconvolution of the parameter group $k/\phi_g^n$. In essence, this issue means that whilst accurate estimates of the compressive and shear yield stress curves can be gained, the individual parameters $k$ and $\phi_g$ are less accurately estimated. In the case of equilibrium height data $h_\infty$, it is necessary to undertake a significant number of tests under variable column height and radii to provide a dense enough dataset for accurate estimation. Prior knowledge of the suspension gel point $\phi_g$ and accurate measurement of $h_\infty$ are required to generate accurate estimates of the shear and compressive rheology. These results indicate that wall adhesion effects can be routinely corrected for in equilibrium batch settling tests, yielding accurate estimates of the shear and compressive rheology of strong colloidal gels. The important case of centrifugal batch settling will be considered in a subsequent paper wherein it will be shown that the errors due to adhesion are normally much less significant, in part because then most of the gel is not near the gel-point and second because $h_\infty/R$ decreases with increasing acceleration.

\section{Acknowledgements}
\label{sec:acknow}
This work was conducted as part of AMIRA P266G ``Improving Thickener Technology'' project, supported by the following companies: Alcoa World Alumina, Anglo American, BASF, Bateman Engineering, Cytec Australia Holdings, Delkor, FL Smidth Minerals, Kemira, MMG, Newcrest, Outotec, Rusal, Senmin, Shell Energy Canada, Teck Resources, Total E\&P Canada and WesTech. The authors are indebted to Jon Halewood for undertaking the flocculation, rheological and solid volume fraction profile measurements, Andrew Chryss for design of rheological measurements.

\end{document}